\documentclass[nofootinbib,aps,prd,preprint,showpacs,superscriptaddress,floatfix]{revtex4-1}
\usepackage{graphicx,graphics}
\usepackage{dcolumn}
\usepackage{amsmath,amssymb,amsfonts}
\usepackage{latexsym,verbatim}
\usepackage{bm}
\usepackage{color}
\usepackage[breaklinks=true,colorlinks,citecolor=blue,linkcolor=blue,urlcolor=blue]{hyperref}
\usepackage{tabularx}

\newcommand{\be}{\begin{equation}}
\newcommand{\ee}{\end{equation}}
\newcommand{\bea}{\begin{eqnarray}}
\newcommand{\eea}{\end{eqnarray}}

\newcommand{\RNumb}[1]{\textbf{\uppercase\expandafter{\romannumeral #1\relax}}}

\newcommand{\eps}{\varepsilon}
\newcommand{\Tr}{\mathop{\mathrm{Tr}}}

\newcommand{\m}{\text{m}}
\newcommand{\gen}{\text{gen}}

\newcommand{\extgen}{^\text{ext}_\text{gen}}

\newcommand{\ba}{\mathbf{a}}
\newcommand{\bb}{\mathbf{b}}

\usepackage{soul}
\usepackage{comment}
\usepackage{xcolor}

\definecolor{darkblue}{rgb}{0,0,1}

\definecolor{dgreen}{rgb}{0,0.6,0}

\definecolor{darkraspberry}{rgb}{0.9,0.,0.3}

\definecolor{aquamarine}{rgb}{0.8,0.0,0.8}

\definecolor{ddgreen}{rgb}{0,0.8,0}

\usepackage{comment}

\begin{document}

\title{Black Holes, Cavities and Blinking Islands}

\author{Dmitry S. Ageev}
\email{ageev@mi-ras.ru}
\affiliation{Steklov Mathematical Institute, Russian Academy of Sciences, Gubkin str. 8, 119991 Moscow, Russia}
\affiliation{Institute for Theoretical and Mathematical Physics, Lomonosov Moscow State University, 119991 Moscow, Russia}

\author{Irina Ya. Aref'eva}
\email{arefeva@mi-ras.ru}
\affiliation{Steklov Mathematical Institute, Russian Academy of Sciences, Gubkin str. 8, 119991 Moscow, Russia}

\author{Timofei A.  Rusalev}
\email{rusalev@mi-ras.ru}
\affiliation{Steklov Mathematical Institute, Russian Academy of Sciences, Gubkin str. 8, 119991 Moscow, Russia}

\date{\today}

\begin{abstract}
Placing a black hole in a cavity provides a natural framework for exploring gravitational scales, thermodynamic instabilities, and effective gravity theories. In this paper, we examine the evolution of entanglement entropy and entanglement islands in a two-sided extension of the Schwarzschild black hole in a cavity. By introducing a reflecting boundary in the eternal black hole exteriors, we regulate the infrared modes of Hawking radiation, finding that the entanglement entropy eventually saturates at a constant value. This value can be lower than the black hole’s thermodynamic entropy, thus avoiding the Page formulation of the information paradox. Regarding entanglement islands, we identify a universal effect induced by the boundary, which we term the ``blinking island''  -- where the entanglement island temporarily disappears, resulting in a short-time information paradox.
\end{abstract}

\maketitle

\section{Introduction}

Black holes and their radiation are mysterious objects that still enigma us, and Hawking radiation is surely one of them \cite{Hawking:1975vcx, Hawking:1976ra}. Recently, much attention has been drawn to the reformulation of the information paradox, or rather its replacement with the question of entanglement evolution in the form of the Page curve  \cite{Page:1993wv, Page:2013dx}. This means, that after the characteristic time called Page time entanglement entropy of radiation exceeds thermodynamic entropy indicating non-unitary evolution \cite{Page:1993wv, Page:2013dx}. 

It  has been  shown how the entanglement island mechanism \cite{Penington:2019npb, Almheiri:2019psf, Almheiri:2019hni} and the replica wormhole~\cite{Penington:2019kki, Almheiri:2019qdq} stop the growth of entanglement after Page time. The entanglement island has been studied in the setups of two-dimensional gravity~\cite{Penington:2019npb, Almheiri:2019psf, Almheiri:2019hni, Almheiri:2019yqk, Penington:2019kki, Almheiri:2019qdq, Chen:2020uac, Almheiri:2020cfm, Chen:2020hmv}, boundary CFT~\cite{Rozali:2019day, Sully:2020pza, Geng:2021iyq, Hollowood:2021wkw, Ageev:2021ipd, Geng:2021mic, Suzuki:2022xwv, Hu:2022ymx, Hu:2022zgy, BasakKumar:2022stg, Afrasiar:2022ebi, Geng:2022dua} and moving mirror models \cite{Davies:1976hi, Good:2016atu, Chen:2017lum, Good:2019tnf, Akal:2020twv, Kawabata:2021hac, Reyes:2021npy, Akal:2021foz, Fernandez-Silvestre:2021ghq, Akal:2022qei}.   

Although the rigorous argument for the existence of entanglement islands is made in lower-dimensional models, in higher-dimensional backgrounds  it is possible to use the s-wave approximation and study them as well under reasonable assumptions. For the case of Schwarzschild black hole solution this has been done in the paper  by Hashimoto, Iizuka and Matsuo \cite{Hashimoto:2020cas}  generalized and discussed in different contexts recently \cite{Ageev:2022qxv, Ageev:2023mzu, Alishahiha:2020qza, Ling:2020laa, Matsuo:2020ypv, Karananas:2020fwx, Wang:2021woy, Kim:2021gzd, Lu:2021gmv, Yu:2021cgi, Ahn:2021chg, Cao:2021ujs, Azarnia:2021uch, Arefeva:2021kfx, He:2021mst, Yu:2021rfg, Arefeva:2022cam, Gan:2022jay, Yu:2022xlh, Piao:2023vgm, Azarnia:2022kmp, Anand:2022mla, Ageev:2022hqc}.

Originally the Page curve is associated with the evaporating black hole setup \cite{Page:1993wv, Page:2013dx}. However, there are others canonical setups to test entanglement dynamics of Hawking radiation and island mechanism that involve the eternal black hole~-- namely the entanglement entropy of infinite regions ``collecting'' Hawking radiation located in different exteriors.

 In this paper, we examine the effects of confining Hawking radiation in each (left and right) exterior with the  perfectly reflecting boundary.

Our motivation for such a study is twofold. The study of the black hole in a cavity introduces a new scale ``regulating'' spatial infinity and clarifies issues concerning the black hole stability, nucleation, etc. \cite{Hawking:1976de, York:1986it,Gross:1982cv}. Black holes in a reflecting box are stable or unstable (heat capacity is positive or negative, respectively) depending on the location of the boundary, with stability achieved when the location is close enough \cite{Hawking:1976de, York:1986it}. The regulation of the infrared mode in entanglement entropy in the context of entanglement islands has been considered in our recent papers \cite{Ageev:2022qxv,Ageev:2023mzu}. Introduction of the boundaries is a reasonable and simple way to regulate infrared issues which can be met in the entanglement entropy of massless fields \cite{Yazdi:2016cxn}. Such a setup may also be useful for studying multi-horizon spacetimes, like the Schwarzschild–de Sitter black hole, where introducing boundaries allows separation of thermal radiation from black hole and cosmological horizons, which have different temperatures \cite{Gibbons:1977mu, Saida:2009ss, Ma:2016arz}. 
Our first motivation is to address this aspect. The second is to find a simple, convenient model in which the island mechanism fails to resolve the Page formulation of the information paradox. We find that a black hole in a cavity exemplifies this case. Introducing a cavity causes the island solution to disappear temporarily—a phenomenon we term the ``blinking island''. For a reflecting boundary placed sufficiently close to the horizon, the entanglement entropy remains below the thermodynamic entropy, thus avoiding issues related to the information paradox.
 
After examining the eternal black hole setup where both exteriors include boundaries, we modify this configuration so that only one exterior region is equipped with a boundary. This adjustment enhances entanglement dynamics for regions contained within a single (left or right) wedge. To our knowledge, such a geometry has not been previously explored. For the double-boundary setup, interpretation in terms of a thermofield double state is straightforward; however, for the single-boundary case, a complete description remains unclear. We interpret this geometry as involving two pairs of distant black holes (one confined by the reflecting boundary), connected by an Einstein–Rosen bridge \cite{Maldacena:2013xja}. When the entangling region (without considering the island mechanism) lies in a single wedge within the double-boundary geometry, the entanglement entropy shows no time dependence. In contrast, with a boundary in only one exterior, we observe a non-trivial dynamical pattern for entanglement. By analogy, we also numerically find asymmetric blinking island solutions in single-boundary geometry, where time-dependent entanglement entropy corresponds non-unitary evolution.

The paper is organized as follows. In Sec. \ref{sec:setup}, we set up the notation, present our path-integral geometries, and review basic formulae from BCFT. Sec. \ref{sec:symm} is devoted to the dynamics of entanglement entropy in the double-boundary geometry and the study of the corresponding blinking island solution. In Sec. \ref{sec:single}, we similarly examine entanglement and islands in the single-boundary geometry. Finally, Sec. \ref{sec:DMR} presents our conclusions.

\section{Setup}\label{sec:setup}

In this section we remind basics of entanglement entropy calculation in two-dimensional BCFT, then describe and generalize Hashimoto-Iizuka-Matsuo \cite{Hashimoto:2020cas} setup    concerning entanglement island study in s-wave approximation. We describe two-dimensional geometry corresponding to eternal black hole containing the reflecting boundaries and conformal mappings neccesary to calculate the entanglement entropy.

\subsection{Entanglement entropy of matter in BCFT$_2$}

Let us start with a brief reminder of the two-dimensional Euclidean boundary conformal field  theory (BCFT) and calculation of the entanglement entropy in such a theory. The simplest 2d BCFT corresponds to the theory defined on upper half-plane (UHP), i.e. evolving on the line. The coordinate~$x_1$ corresponds to the Euclidean time, while $x_2$ is the spatial coordinate
\be
ds^2 = dx^2_1+dx^2_2 = dz d\bar{z}, \quad z = x_1 + i x_2, \quad x_1 \in (-\infty, \infty), \quad x_2 \geq 0,
\ee
We will be interested in the  study of region $R$   consisting  of single intervals union
\be\label{eq:region}
R = [z_{a_1}, z_{b_1}] \cup \ldots \cup [z_{a_n}, z_{b_n}],
\ee
and its entanglement entropy
\be
S_\m(R) = -\Tr (\rho_R \log \rho_R),
\ee
where $\rho_R$ is reduced density matrix for region $R$. Entanglement entropy could be obtained by the replica trick \cite{Calabrese:2004eu, Calabrese:2009qy} as
\be\label{eq:entropy-from-replica-trick}
S_\m (R) = - \lim_{n \to 1} \partial_n \left(\Tr \rho^n_R\right).
\ee 
The trace in \eqref{eq:entropy-from-replica-trick} is expressed as a correlation function of twist operators inserted at the bulk endpoints of intervals of region $R$ that do not belong to the boundary $x_2=0$
\be\label{eq:correlator}
\Tr \rho^n_R = \langle \phi (z_{a_1}, \bar{z}_{a_1}) \phi (z_{b_1}, \bar{z}_{b_1}) \ldots \phi (z_{a_n}, \bar{z}_{a_n}) \phi (z_{b_n}, \bar{z}_{b_n})  \rangle_{\text{UHP}}.
\ee
If the endpoint of the interval is on the boundary, for example $R = [0, z_1]$, then the trace corresponding to such interval is given by  $\Tr \rho^n_R = \langle \phi (z_1, \bar{z}_1) \rangle_{\text{UHP}}$. The twist operators are primary operators with conformal dimensions $h_n=\bar{h}_n = c/24 (n-1/n)$, where $c$ is the central charge \cite{Calabrese:2009qy}.
\\

We consider the particular BCFT$_2$ given by $c$ copies of two-dimensional free massless Dirac fermions. The Dirac field is given by the doublet of the left and right moving components
\be
\psi (x_1, x_2) = \bigg(
\begin{array}{c}\psi_1(x_1, x_2) \\ \psi_2(x_1, x_2) \\ \end{array}
\bigg) = \bigg(
\begin{array}{c}\psi_1(z) \\ \psi_2(\bar{z}) \\ \end{array}
\bigg).
\ee
We choose the boundary conditions  preserving the conformal invariance of the theory and corresponding to the vanishing of the energy and momentum flow through the boundary \cite{Cardy:1984bb, Cardy:1986gw, Cardy:1989ir}. There are two explicit boundary conditions that satisfy this requirement for massless free Dirac fermions on UHP \cite{Mintchev:2020uom, Rottoli:2022plr}:
\be\label{eq:BC-vector-phase}
\psi_1 (x_1, 0) = e^{i \alpha_v} \psi_2 (x_1, 0), \qquad \alpha_v \in [0, 2\pi), \, \, x_1 \in (-\infty, \infty),
\ee
or
\be\label{eq:BC-axial-phase}
\psi_1 (x_1, 0) = e^{-i \alpha_a} \psi^{*}_2 (x_1, 0), \qquad \alpha_a \in [0, 2\pi), \, \, x_1 \in (-\infty, \infty).
\ee
These boundary conditions correspond to the transition of the left moving component to the right one, which can be interpreted as perfectly reflecting boundary conditions. The entanglement entropy of a region \eqref{eq:region} is given by \cite{Kruthoff:2021vgv, Rottoli:2022plr} 
\be
\begin{aligned}
S (R) = & \frac{1}{3} \sum^{n}_{i,j=1}\log |z_{a_i} - z_{b_j}| -\frac{1}{3}  \sum^{n}_{i<j} \log  |z_{a_i} - z_{a_j}| |z_{b_i} - z_{b_j}|-n \log \varepsilon  \\ + &\frac{1}{6} \sum^{n}_{i,j=1}\log |z_{a_i}-\bar{z}_{a_j}| |z_{b_i}-\bar{z}_{b_j}| -\frac{1}{6} \sum^{n}_{i,j=1}\log |z_{a_i}-\bar{z}_{b_j}| |z_{b_i}-\bar{z}_{a_j}|,
    \end{aligned}
    \label{eq:entropy-for-Dirac-fermions-UHP}
\ee
where $\varepsilon$ is UV cutoff. The entropy \eqref{eq:entropy-for-Dirac-fermions-UHP} does not depend on the choice of the boundary condition \eqref{eq:BC-vector-phase}  or \eqref{eq:BC-axial-phase} and on the phase $\alpha_{v, a}$ \cite{Mintchev:2020uom, Rottoli:2022plr}.
\\

If the state (corresponding to some geometry $\Omega$) of BCFT$_2$ is related with the UHP, then  under conformal transformation 
\be\label{eq:conf-transf-general}
z: \Omega \to \text{UHP}, \quad  z = z(w), \quad \bar{z} = \bar{z} (\bar{w}).
\ee
 correlation function of twist operators  transforms as
\be
\begin{aligned}
\langle \phi (w_1, \bar{w}_1) \ldots \phi (w_m, \bar{w}_m) \rangle_{\Omega} =  \prod\limits_{j = 1}^m & \left( \frac{dz}{dw}\right)^{h_n}\Big|_{w=w_j} \left( \frac{d\bar{z}}{d\bar{w}}\right)^{\bar{h}_n}\Big|_{\bar{w}=\bar{w}_j}   \\ \times & \langle \phi (z_1, \bar{z}_1) \ldots \phi (z_m, \bar{z}_m) \rangle_{\text{UHP}}. 
    \end{aligned}
   \label{eq:conf}
\ee
Entanglement entropies in flat $ds^2 = dw d\bar{w}$ and curved $ds^2 = e^{2 \rho (w, \bar{w}) } dw d\bar{w}$ two-dimensional spacetimes are related by Weyl transformation $ds^2 \to e^{2\rho} ds^2$
\be \label{eq:weyl}
S_\m\Big|_{ds^2 = e^{2 \rho (w, \bar{w}) } dw d\bar{w}}= S_\m\Big|_{ds^2 =  dw d\bar{w}} + \frac{c}{6} \sum_{i=1}^m \log e^{\rho (w_i, \bar{w}_i) }.
\ee

The calculation of entanglement entropy $S_{\m}$ of matter theory in a higher-dimensional curved spacetime is an extremely challenging problem. For the study of entanglement islands in \cite{Hashimoto:2020cas} one proposed  to consider the s-wave approximation to overcome difficulties related to higher dimensional geometries (Schwarzschild black hole). So following~\cite{Hashimoto:2020cas} for spherically symmetric geometries of the form
\be
ds^2 = e^{2 \rho (w, \bar{w}) } dw d\bar{w}+r^2d\Omega_d^2
\ee
in what follows we  effectively neglect the spherical part of the metric and consider two-dimensional (B)CFT on the rest of the spacetime
\be 
ds^2 = e^{2 \rho (w, \bar{w}) } dw d\bar{w},
\ee 
assuming that the obtained results captures all main features of dynamics on the higher-dimensional original background.

\subsection{Geometry, introduction of the boundaries and path-integral}

Our main geometry to study is given by the metric of the four-dimensional Schwarzschild black hole
\be
    ds^2 = -f(r)dt^2 + \frac{dr^2}{f(r)} + r^2 d\Omega_2^2, \qquad f(r) = 1 - \frac{r_h}{r},
    \label{eq:Sch_metric}
\ee
at $r > r_h$, $-\infty < t < \infty$, where $r_h = 2 G M$ denotes the black hole horizon, $M$ is the mass of the black hole, $G$ is the gravitational constant and $d\Omega_2^2$ is the angular part of the metric. As we already mentioned in the previous section, in what follows we consider only the two-dimensional part of the metric~\eqref{eq:Sch_metric} relying on the s-wave approximation. Within its framework omitting the angular variables, introducing Kruskal coordinates
\be
    U = -\frac{1}{\kappa_h}\,e^{-\kappa_h(t - r_*(r))}, \quad V = \frac{1}{\kappa_h}\,e^{\kappa_h(t + r_*(r))},\,\,\,\,\kappa_h = 1/2r_h,
    \label{eq:right_wedge_Krusk}
\ee
 we study two-dimensional metric of the form
\be
    ds^2 = -e^{2 \rho(r)} dU dV, \quad
    e^{2 \rho(r)} = f(r) e^{-2 \kappa_h r_{*} (r)}.
    \label{eq:krusk-metr}
\ee
where $r_*(r)$ is tortoise coordinate given by $r_*(r) = r + r_h\log |r - r_h|/r_h$.

The transformation between Schwarzschild and Kruskal coordinates \eqref{eq:right_wedge_Krusk} can be analytically extended to the region $-\infty < U,V < \infty$ with the constraint $U V < 1 / \kappa^2_h$, which is associated with the physical singularities of the past and the future. 
It is convenient now to relate Kruskal coordinates to  timelike $T$ and spacelike $X$ variables
\bea\label{eq:time-spacelike-Kruskal}
U =  T-X, \qquad V  =  T+X.
\eea
 expressed in terms of Schwarzschild coordinates as 
\bea\label{eq:timespacelike-to-Scwh-Lor}
T = \pm \frac{e^{\kappa_h r_{*} (r)}}{\kappa_h} \sinh \kappa_h t, \qquad
X = \pm \frac{e^{\kappa_h r_{*} (r)}}{\kappa_h} \cosh \kappa_h t,
\eea
where the upper (lower) sign corresponds to the right (left) wedge.

We  Wick rotate  Kruskal time $T = -i {\cal T}$ and this also defines  the Euclidean Schwarzschild time $\tau = i t$  periodic with a period of $2 \pi /\kappa_h$. The coordinate  transformation \eqref{eq:timespacelike-to-Scwh-Lor} now takes the form
\bea
{\cal{T}} = \pm \frac{e^{\kappa_h r_{*} (r)}}{\kappa_h} \sin \kappa_h \tau, \qquad
X = \pm \frac{e^{\kappa_h r_{*} (r)}}{\kappa_h} \cos \kappa_h \tau. 
\eea
and  left (right) half-plane $X<0$ ($X>0$) corresponds to the left (right) wedge of Lorentzian black hole. To calculate the entanglement entropy we start with some  the Euclidean geometry defining path integral and state, then calculate necessary correlation functions and get back to Lorentzian signature in the end.

$\,$

From the identity \bea\label{eq:euclidean-constant-surfaces}
X^2+{\cal T}^2 = \frac{e^{2\kappa_h r_{*} (r)}}{\kappa^2_h}, \qquad \tan \kappa_h \tau = \frac{{\cal T}}{X},
\eea
it is straightforward to see that  in the (${\cal T}$, $X$) plane curves $r = \text{const}$ correspond to circles centered at the point  ${\cal T}=X=0$ and $\tau = \text{const}$ to the straight lines passing through this point. We consider the exterior of the Euclidean black hole, i.e. $r\geq r_h$, and the origin of the (${\cal T}$, $X$) plane corresponds to $r = r_h$. 

\begin{figure}[t!]\centering
    \includegraphics[width=1.\textwidth]{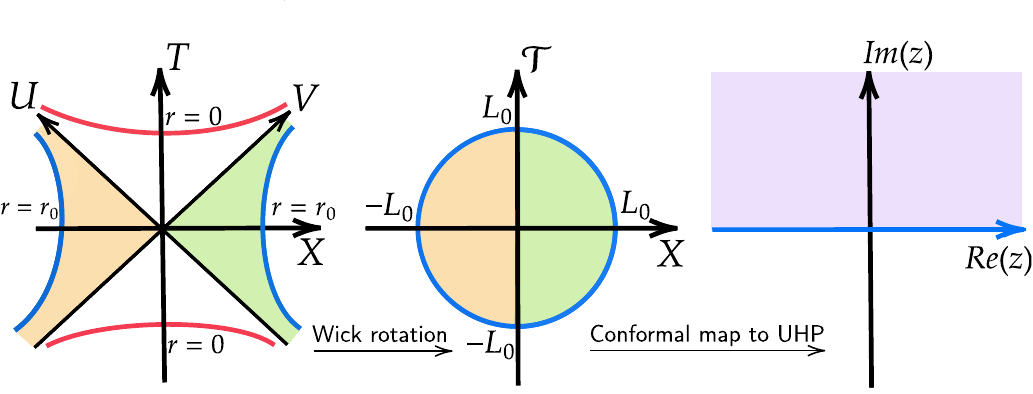}
	\caption{Lorentzian (\textbf{left}) and Euclidean (\textbf{center}) Kruskal diagram of a two-sided Schwarzschild black hole in the presence of two symmetric boundaries at $r=r_0$ drawn in blue. The Euclidean Kruskal diagram corresponds to the interior of the disk, that maps conformally to the upper half-plane (\textbf{right}).}
  \label{fig:disc-setup}
\end{figure}

We introduce a complex coordinates 
\be\label{eq:complex-coordinate-omega}
w = X+ i {\cal T}, \qquad \bar{w} = X- i{\cal T},
\ee
and in this coordinates the Euclidean version of the metric \eqref{eq:krusk-metr}   has the form
\be\label{eq:euclidean-metric-complex-coordinates}
ds^2 =   e^{2 \rho (w,\bar{w})} dw d\bar{w}, \qquad  e^{2 \rho (w,\bar{w})} = \frac{W( e^{-1} \kappa^2_h w \bar{w} )}{\kappa^2_h w \bar{w} \left[1+W(e^{-1} \kappa^2_h w  \bar{w} )\right]},
\ee
where $W(x)$ is Lambert W function. Total geometry given by \eqref{eq:euclidean-metric-complex-coordinates} (i.e. at $-\infty < {\cal T}, X < \infty$) is just the complex plane endowed with the non-trivial metric.
\\

\begin{figure}[t!]\centering
    \includegraphics[width=1.\textwidth]{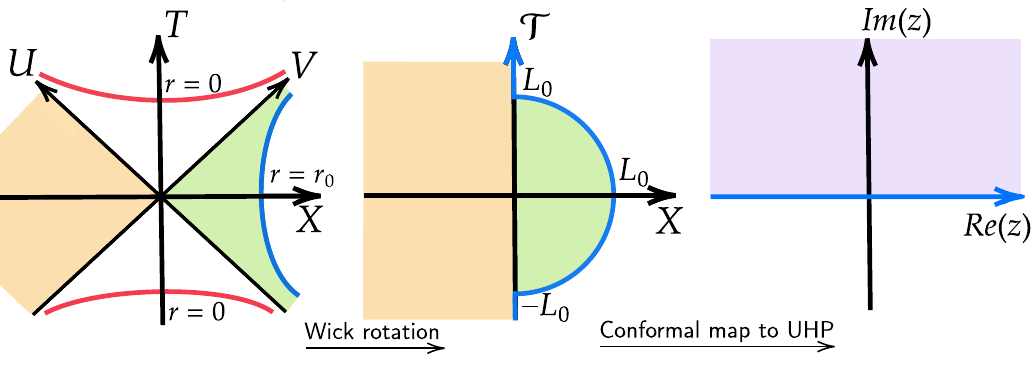}
	\caption{Lorentzian (\textbf{left}) and Euclidean (\textbf{center}) Kruskal diagram of a two-sided Schwarzschild black hole in the presense of boundary, drawn in blue, at radius $r=r_0$ only in the right wedge. The Euclidean Kruskal digram corresponds to the union of the interior of the half-disk and the left half-plane, that maps conformally to the upper half-plane (\textbf{right}).}
 \label{fig:zaslonka-setup}
\end{figure}

Now we  introduce spherically symmetric boundary at the radial coordinate $r~=~r_0$, where $r_0> r_h$, in the analytically extended Schwarzschild geometry. Quantum field (Dirac fermions)  now are  restricted by   the boundary at  $r < r_0$ via reflecting boundary condition being imposed. We consider two ways to introduce a boundary:
\begin{itemize}
    \item Two  boundaries both in the right and left wedges are located at the same radial coordinate. The Euclidean geometry in the plane $({\cal T}, X)$  corresponds to the interior of a disk with the radius $L_0 = e^{\kappa_h r_{*} (r_0)}/\kappa_h$, see Fig.\ref{fig:disc-setup}
    \item The boundary is located only in the one of the wedges, for examples in the right one for concreteness. The Euclidean geometry in the plane $({\cal T}, X)$  corresponds to the union of left half-plane $X<0$ and the interior of a semi-disk in the right half-plane $X>0$ with the radius $L_0 = e^{\kappa_h r_{*} (r_0)}/\kappa_h$, see Fig.\ref{fig:zaslonka-setup}
\end{itemize}

From the path-integral point of view the wavefunctional of our interest is given by the  integration over the geometry part with ${\cal T}<0$, i.e. over the lower half of the disk   from Fig.\ref{fig:disc-setup} (for the case of symmetric boundaries). For the single boundary located in the right wedge path integral defining wavefunction is performed over the union of the lower left quadrant and the lower half of the half-disk (${\cal T}<0$ region in Fig.\ref{fig:zaslonka-setup}).

$\,$

As our ultimate goal is to study the entanglement entropy the next step is to perform conformal mapping of the Euclidean geometry given by the central picture in Fig.\ref{fig:disc-setup} or  Fig.\ref{fig:zaslonka-setup} to the upper half-plane with non-trivial curved metric  and then  implement Weyl transformation to the flat metric upper half-plane. In this way we obtain explicit form of the entanglement entropy given by the composition of transformation rules for conformal mappings \eqref{eq:conf} and  Weyl transformations \eqref{eq:weyl}. The map from the interior of the disk with radius $L_0$ to the UHP is defined by 
\be \label{eq:symm-map}
z = i \frac{L_0+w}{L_0-w}, \qquad \bar{z} = -i \frac{L_0+\bar{w}}{L_0-\bar{w}}.
\ee
In its turn the map from the union of left half-plane $X<0$ and the interior of a semi-disk in the right half-plane $X>0$ (see Fig.\ref{fig:zaslonka-setup}) with the radius $L_0$ to the UHP is given by\footnote{We assume taking the root of the third degree of a complex number $\eta = \rho e^{i\varphi}$ implies the principal branch that corresponds to $\eta^{1/3} = \rho^{1/3} e^{i\varphi/3}$ (see Appendix \ref{appendix-A} for details).}
\be\label{eq:conform-transformation-one-boundary}
z = e^{2\pi i/ 3} \left( \frac{L_0+i w}{L_0-i w}\right)^{\frac{2}{3}}, \qquad \bar{z} = e^{-2\pi i/ 3} \left( \frac{L_0-i \bar{w}}{L_0+i \bar{w}}\right)^{\frac{2}{3}}.
\ee

Finally let us briefly comment on the possible interpretations of the geometries under consideration. For the boundaries located symmetrically in the left and right wedges one can refer to the thermofield interpretation of the eternal black hole described by the entangled thermofield double state at $t=0$
\be\label{eq:thermofield-double-state}
| \Psi \rangle = \sum_n e^{-\beta E_n/2} |n\rangle_L |n\rangle_R,
\ee
where $|n \rangle_{L, R}$ are the eigenstates with energy $E_n$ of the matter theory Hamiltonian in the left/right wedges, $\beta = \kappa_h/2\pi$ is the inverse temperature of the black hole. 

 In this interpretation, the time evolution is upward in both left and right wedges with Hamiltonian
\be\label{eq:sum-hamiltonians}
H_{tot} = H_L + H_R.
\ee
While the state with symmetric boundaries is quite clear from thermofield intepretation point of view, a similar interpretation of the state for single boundary geometry does not seem to be valid due to the lack of symmetry of the right and left wedges.

\subsection{Generalized entropy functional}

Recently, it was shown that the expected behavior of the Page curve emerges from the island proposal~\cite{ Almheiri:2019hni, Penington:2019npb, Almheiri:2019qdq}. The reduced density matrix of Hawking radiation collected in~$R$ is defined by tracing out the states in the complement region~$\overline{R}$, which includes the black hole interior. The island mechanism prescribes that the states in some regions~${I \subset \overline{R}}$, called entanglement islands, are to be excluded from tracing out.

The island contribution can be taken into account via the generalized entropy functional defined as~\cite{Penington:2019kki, Almheiri:2019qdq}
\be
    S_\gen[I, R] = \frac{\operatorname{Area}(\partial I)}{4 G} + S_\m(R \cup I).
    \label{eq:gen_functional}
\ee
Here $\partial I$ denotes the boundary of the entanglement island, $G$ is Newton's constant, and $S_\m$ is the entanglement entropy of conformal matter. One should extremize this functional over all possible island configurations
\be\label{eq:extremization-procedure}
    S\extgen[I, R] = \underset{\partial I}{\operatorname{ext}}\,\Big\{S_\gen[I, R]\Big\},
\ee
and then choose the minimal one
\be\label{eq:island-formula}
    S(R) = \underset{\partial I}{\text{min}}\,\Big\{S\extgen[I, R]\Big\}.
\ee
Note that the entropy of conformal matter $S_\m$ is proportional to the number of fields~$c$, while it is believed that the area term in the case of a nontrivial island configuration is proportional to~$r^2_h/G$. Since the area term is the leading classical contribution and the entropy of conformal matter is a quantum correction, we impose the constraint $c G/r^2_h \ll~1$ (or, equivalently, $c G \kappa^2_h \ll 1$). Also, this condition allows us to neglect the backreaction effect of matter fields on geometry.

\section{Symmetrical boundaries in both exteriors}\label{sec:symm}
\begin{figure}[h!]\centering
    \includegraphics[width=0.915\textwidth]{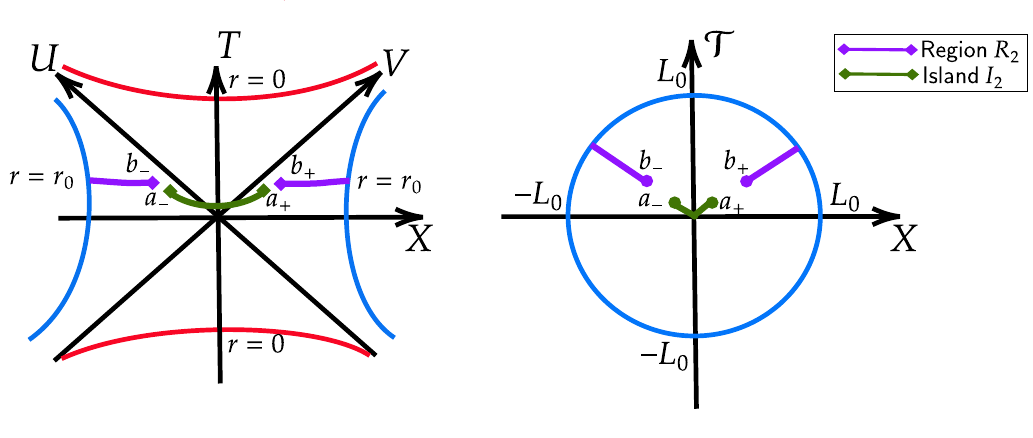}
	\caption{The schematic depicture of the region $R_2$ (magenta) in: two-boundary Lorentzian geometry (\textbf{left}) and its Euclidean version (\textbf{right}).} \label{fig:region-symmm}
\end{figure}

Now let us  study the first proposed in the setup geometries -- black hole bounded by two symmetric reflecting boundaries. We calculate the entanglement entropy of the region $R_2$, which is the union of two intervals located  between bulk points $\bb_\pm$ and the boundary in the corresponding left/right wedge (magenta curves in Fig.\ref{fig:region-symmm}). In the island phase we assume the inclusion of the island region $I_2$ given by the ``interval'' between two points $\ba_-$ and $\ba_+$ in different wedges (green curves in Fig.\ref{fig:region-symmm}) 

\subsection{Without island}

First apply the strategy of entanglement entropy calculation described in the previous section to the evolution of  entanglement entropy  \eqref{eq:entropy-for-Dirac-fermions-UHP}--\eqref{eq:weyl}  with  conformal mapping~\eqref{eq:symm-map} and Weyl transformation of the resulting curved UHP to the flat one. Here we study the entanglement entropy without island phase and explicitly choose the points $\bb_\pm$ as 
\be\label{eq:points-for-our-region}
\bb_+ = \left(r_b, t_b \right), \qquad \bb_- = \left(r_b, -t_b \right).
\ee 
\begin{figure}[h!]\centering
    \includegraphics[width=0.55\textwidth]{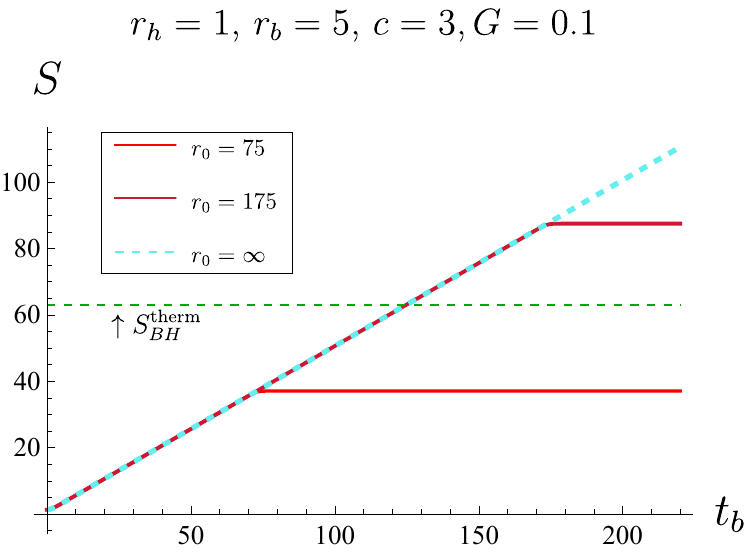}
	\caption{Time dependence of entanglement entropy without an island \eqref{eq-disc:entropynoisland1} for different boundary positions. The entanglement entropy initially increases monotonically the same way as in the case without boundaries ($r_0 = \infty$), and approximately at the time \eqref{eq:time-tb-1} the entropy reaches saturation \eqref{eq:disc-saturation-value}. The  entropy saturation time increases with the increase of boundary location $r_0$. For a sufficiently far boundary, the entanglement entropy exceeds the thermodynamic entropy of a black hole $S^{\text{therm}}_{BH}$ \eqref{eq:thermodynamic-entropy-of-black-hole} (dashed green line).}
 \label{fig:entropy-disc-no-island}
\end{figure}

With this choice of points after some algebra one writes down the time-dependent entanglement entropy of $R_2$ as
\be\label{eq-disc:entropynoisland1}
S(R_2) = \frac{c}{6} \log \left( \frac{4f(b)  \cosh^2 \kappa_h t_b}{\kappa^2_h \eps^2}\right) + \frac{c}{6} \log\left( \frac{2 \sinh^2 \kappa_h (r_{*} (r_0)-r_{*} (r_b))}{\cosh 2\kappa_h (r_{*} (r_0)-r_{*} (r_b))+\cosh 2\kappa_h t_b}\right).
\ee
Note that in the limit $r_0 \to \infty$ the second term of \eqref{eq-disc:entropynoisland1} tends to zero, so the entropy without boundary (the first term) is restored.
At large times
\be\label{eq:time-tb-1}
t_b \gg t^1_b, \qquad t^1_b \equiv r_{*} (r_0)-r_{*} (r_b),
\ee
the entropy \eqref{eq-disc:entropynoisland1} saturates at the value
\be\label{eq:disc-saturation-value}
S(R_2) = \frac{c}{6} \log \left( \frac{4f(b)  \sinh^2 \kappa_h (r_{*} (r_0)-r_{*} (r_b))}{\kappa^2_h \eps^2}\right).
\ee
The time dependence of the entanglement entropy \eqref{eq-disc:entropynoisland1} for different positions of the boundary $r_0$ and the comparison with the entanglement entropy for the case without boundaries (i.e. at $r_0 \to \infty$) are shown in Fig. \ref{fig:entropy-disc-no-island}.

The time $t^1_b$ \eqref{eq:time-tb-1}, until which the entanglement entropy \eqref{eq-disc:entropynoisland1} increases monotonically, depends on the location of the boundary $r_0$. For a sufficiently large $r_0$, the entropy~\eqref{eq-disc:entropynoisland1} at some time $t_b < t^1_b$   exceeds  the thermodynamic entropy of the black hole
\be\label{eq:thermodynamic-entropy-of-black-hole}
S^{\text{therm}}_{BH} = \frac{2\pi r^2_h}{G},
\ee
which is equal to twice the Bekenstein-Hawking entropy.\footnote{The thermodynamic entropy of a black hole does not change when a boundary is introduced \cite{York:1986it}.} This leads to a violation of the unitarity of the evolution of the closed system ``black hole + Hawking radiation'', which we interpret as an information paradox for a two-sided black hole with symmetric boundaries.

\subsection{Entanglement entropy for AdS-Schwarzschild black hole}

In the previous subsection we found that for an asymptotically flat Schwarzschild black hole with a reflecting boundary the presence of the information paradox depends on the location of the boundary. The information paradox holds for a sufficiently distant boundary (for a ``small'' black hole) and is absent for a sufficiently close one (for a``large'' black hole). This is consistent with the fact that black holes in a reflecting box are stable or unstable (in the sense of positive or negative heat capacity, respectively) depending on the location of the boundary, with stability being achieved at a sufficiently close location~\cite{Hawking:1976de, York:1986it}. However, in this paper we do not explicitly verify that the information paradox is associated with thermodynamic instability.

At least the same thermodynamic behavior is observed in the AdS-Schwarzschild black hole without flat baths attached to the conformal boundary.\footnote{We thank an anonymous referee for highlighting this issue.} Indeed, for this case, there is a well-known classification into small and large black holes.

\begin{figure}[h!]\centering
    \includegraphics[width=0.55\textwidth]{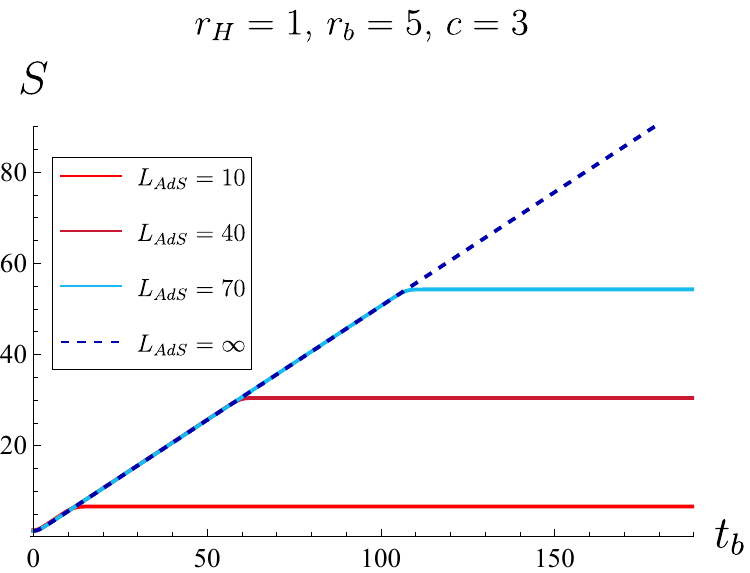}
	\caption{Time dependence of entanglement entropy without an island, as given by \eqref{eq-disc:entropynoisland1}, for the AdS-Schwarzschild black hole with different values of $L_{AdS}$. The entanglement entropy initially increases monotonically and eventually saturates. The time to reach saturation increases with $L_{AdS}$. As $L_{AdS} \to \infty$, the entropy grows monotonically, as indicated by the dashed dark blue line.
}
 \label{fig:entropy-disc-no-island-AdS}
\end{figure}

Let us consider the entanglement entropy of radiation in the two-sided geometry of the eternal AdS-Schwarzschild black hole. Specifically, we use the metric in the form \eqref{eq:Sch_metric} with
\be
f(r) = 1 - \frac{r_H}{r} \left(1 + \frac{r_H^2}{L^2_{AdS}} \right) + \frac{r^2}{L^2_{AdS}},
\ee
where $r_H$ is the horizon location and $L_{AdS}$ is the radius of AdS \cite{Hawking:1982dh}. We also define the corresponding surface gravity $\kappa_h = f'(r_H)/2$ and the tortoise coordinate $r_*(r) = \int \frac{dr'}{f(r')}$. In this background, we consider a spacelike region $R$ similar to $R_2$ discussed above, extending from symmetric points $\bb_{\pm}$, as in \eqref{eq:points-for-our-region}, in the left and right wedges to the corresponding conformal boundaries. As before, the radial coordinate $r_b$ of the endpoints $\bb_{\pm}$ satisfies $r_b > r_H$. The entanglement entropy of this region takes the same form as in \eqref{eq-disc:entropynoisland1}, with $f(r)$, $\kappa_h$, and $r_*(r)$ as defined above. The behavior of the entanglement entropy for different values of $L_{AdS}$ is shown in Fig.~\ref{fig:entropy-disc-no-island-AdS}.

Qualitatively, we obtain the same situation as in the case of the asymptotically flat black hole with a boundary, as shown in Fig.~\ref{fig:entropy-disc-no-island}. For a sufficiently ``distant'' conformal boundary (i.e., large $L_{AdS}$ for fixed $r_H$, corresponding to ``small'' black holes), the saturation of the entanglement entropy exceeds the thermodynamic entropy of the black hole. However, for small $L_{AdS}$ (i.e., ``large'' black holes), this does not occur. As $L_{AdS} \to \infty$, the entropy grows monotonically, as expected for an asymptotically flat black hole without a reflecting boundary \cite{Hashimoto:2020cas}. In other words, the information paradox is again understood to depend on the size of the black hole.

Note that there is also a difference between an asymptotically flat black hole with a boundary and an AdS-Schwarzschild without flat baths. Namely, in the former case, when choosing $r_b$ such that $r_b \gg r_h$ (this is what we consider in this paper), we can effectively assume that the spacelike region $R_2$ is located in a domain with weak gravity (``freezing'' of dynamic gravity), since the metric \eqref{eq:Sch_metric} tends to a flat one. In fact, this is not related to the presence or absence of a reflecting boundary, and was used in \cite{Hashimoto:2020cas, Alishahiha:2020qza} within the framework of the s-wave approximation. In other words, in asymptotically flat geometry there is a domain of spacetime that can effectively be considered as ``flat non-dynamical baths". Whereas in AdS-Schwarzschild the curvature is non-zero in any domain of spacetime. In this sense, our setup with an asymptotically flat black hole with reflective boundaries is preferable to AdS-Schwarzschild without flat baths. In the rest of the paper we do not consider AdS-Schwarzschild anymore and postpone a more detailed study of this issue for future research. Note that the study of the radiation entanglement entropy on the background of a planar eternal AdS black hole was carried out in \cite{Geng:2022dua}.

\subsection{Blinking island solution}

In the previous subsection it was shown that depending on the position of the boundary one can observe the exceed of entanglement entropy over black hole entropy. In this subsection we consider the entanglement entropy with a non-trivial island and examine whether this obstructs the excessive entanglement growth thus avoiding non-unitary dynamics i.e. 
\be\label{eq:condition-of-unitarity}
S(R_2) (t_b) < S^{\text{therm}}_{BH}
\ee
is to be satisfied at all values  of time $t_b$.

Due to the symmetric choice of endpoints $\bb_{\pm}$ \eqref{eq:points-for-our-region} of the region $R_2$ and the symmetrical location of boundaries in both wedges, it follows that the endpoints~$\ba_{\pm}$ of island $I_2$ are also symmetric and are given as follows
\be\label{eq:points-for-symmetric-island}
\ba_+ = \left(r_a, t_a \right), \qquad \ba_- = \left(r_a, -t_a \right),
\ee 
where the points $\ba_{+}$/$\ba_{-}$ are located in the right (left) wedge.
The generalized entropy~\eqref{eq:gen_functional} with the symmetric island $I_2$ is
\be
S_\gen[I_2, R_2] = S^{\text{wb}}_{I} (R_2)+S^{\text{b}}_{I} (R_2),
\label{eq-disc:entropyWithisland11}
\ee
where the first term $S^{\text{wb}}_{I} (R_2)$ is independent of radial coordinate $r_0$ of boundary. In the limit  $r_0 \to \infty$ we are left only with this term corresponding  to the generalized entropy for a two-sided black hole with regions extending to spacelike infinity \cite{Hashimoto:2020cas} 
\be
\begin{aligned}
S^{\text{wb}}_{I} (R_2) &=  \frac{2 \pi r^2_a}{G}+\frac{c}{3} \log \left(\frac{4 \sqrt{ f(r_a) f(r_b)} \cosh \kappa_h t_a \cosh \kappa_h t_b}{\kappa^2_h \eps^2} \right) \\ &+ \frac{c}{3} \log \left( \frac{\cosh \kappa_h (r_{*} (r_a)-r_{*} (r_b))-\cosh\kappa_h (t_a-t_b)}{\cosh \kappa_h (r_{*} (r_a)-r_{*} (r_b))+\cosh\kappa_h (t_a+t_b)}\right).
    \end{aligned}
\label{eq-disc:entropywithisland-without-boundary}
\ee
The second term of generalized entropy \eqref{eq-disc:entropyWithisland11} is the contribution  purely due to the  boundary inclusion
\be
\begin{aligned}
&S^{\text{b}}_{I} (R_2) =   \frac{c}{3} \log \left( \frac{\cosh \kappa_h  (2r_{*} (r_0)-r_{*} (r_b)-r_{*} (r_a)) + \cosh \kappa_h (t_a+t_b)}{\cosh \kappa_h  (2r_{*} (r_0)-r_{*} (r_b)-r_{*} (r_a)) -\cosh \kappa_h (t_a-t_b)}\right) \\ & + \frac{c}{6} \log \left( \frac{4 \sinh^2 \kappa_h (r_{*} (r_0)-r_{*} (r_a)) \sinh^2 \kappa_h (r_{*} (r_0)-r_{*} (r_b))}{(\cosh 2\kappa_h  (r_{*} (r_0)-r_{*} (r_a)) + \cosh 2\kappa_h t_a)(\cosh 2\kappa_h  (r_{*} (r_0)-r_{*} (r_b)) + \cosh 2\kappa_h t_b)} \right) .
    \end{aligned}
\label{eq-disc:entropywithisland-boundary}
\ee

In accordance with the prescription of the island formula \eqref{eq:island-formula}, it is necessary to carry out extremization with respect to the radial and time coordinates $(r_a, t_a)$ of the island boundaries in \eqref{eq-disc:entropyWithisland11}, i.e. find the solutions of
\begin{equation}\label{eq:extremization-system-disc}
 \begin{cases}
   \partial_{r_a} S_\gen[I_2, R_2] \left(r_a, t_a, r_b, t_b, r_0, r_h, c, G \right) = 0, 
   \\
   \partial_{t_a} S_\gen[I_2, R_2] \left(r_a, t_a, r_b, t_b, r_0, r_h, c, G \right) = 0.
 \end{cases}
\end{equation}
We consider solutions to the system \eqref{eq:extremization-system-disc}  corresponding to the location of the island boundary $r_a$ near the horizon, i.e. $r_a = r_h + X$, $X/r_h \ll 1$.\footnote{Note that in the case without boundaries there is a solution $r_a \simeq r_b$, $t_a = t_b$ \cite{He:2021mst}. In our case with symmetric boundaries, there is also a similar solution for all times $t_b \gg r_h$, but the area term of~\eqref{eq:gen_functional} for such a solution is tens of times greater than the thermodynamic entropy of the black hole~\eqref{eq:thermodynamic-entropy-of-black-hole} due to $r_b \gg r_h$, and therefore does not help eliminate non-unitary evolution. Therefore, we do not examine this solution in detail.} We also work within the framework of the condition $r_b \gg r_h$ (for the applicability of the s-wave approximation~\cite{Hashimoto:2020cas}) and   also assuming $c G \kappa^2_h \ll 1$.

For fixed parameters $(r_b, r_0, r_h, c, G)$, the solution (or lack thereof) of the system~\eqref{eq:extremization-system-disc} significantly depends on the time $t_b$. There are three different time $t_b$ regimes. Let us study each of them in detail.
\begin{itemize}
    \item \underline{Early time regime}:

    This time regime is determined by the following condition
    \be
        \begin{aligned}
        \cosh \kappa_h (t_a + t_b) &\ll  \cosh \kappa_h (2r_{*} (r_0) - r_{*} (r_{a})-r_{*} (r_{b})), \\ 
        \cosh 2 \kappa_h t_{a} &\ll \cosh 2 \kappa_h (r_{*} (r_0) - r_{*} (r_{a})).
        \end{aligned}
        \label{eq-disc:early-time-regim-condition}
    \ee

    If along with \eqref{eq-disc:early-time-regim-condition} the condition
    \be\label{eq:time-for-existense-of-island-in-early-regim}
    \cosh \kappa_h (t_a + t_b) \gg        \cosh \kappa_h (r_{*} (r_b)-r_{*} (r_a))
    \ee
    is also satisfied, then one can show that there is a non-trivial solution of the system \eqref{eq:extremization-system-disc} 
    \bea\label{eq:early-time-ta}
    t_a &\approx& t_b - \frac{1}{2 \kappa_h} \log\left[1-e^{2\kappa_h (t_b-r_{*}(r_0)+r_{*}(r_b))}\right], \\
    r_a &\approx& r_h+\frac{c^2 G^2 e^{(r_h-r_b)/r_h}}{144 \pi^2 r^2_h (r_b-r_h)}-\frac{c^2 G^2 e^{(t_b+r_h-r_0)/r_h}}{144 \pi^2 r^2_h (r_0-r_h)}. \label{eq:early-time-ra}
    \eea
    In the limit $r_0 \to \infty$ the solution \eqref{eq:early-time-ta}, \eqref{eq:early-time-ra} corresponds to the solution for the case without boundaries \cite{Hashimoto:2020cas}. From the condition that \eqref{eq:early-time-ta} is real, it follows that the solution in the early time regime exists only for
    \be\label{eq:early-time-condition-that-island-exist}
    t_b < t^1_b, \qquad t^1_b = r_{*}(r_0)-r_{*}(r_b).
    \ee
    Recall that time $t^1_b$ approximately corresponds to the moment when the entanglement entropy without an island \eqref{eq-disc:entropynoisland1} reaches saturation.
    
    \item \underline{Intermediate time regime}:
    
    This time regime is determined by the following condition\footnote{Another version of the intermediate regime, when the inequalities \eqref{eq-disc:intermediate-time-regim-condition} are reversed, cannot exist.  From the condition $t_a >  r_{*}(r_0)-r_{*}(r_a)$ and from the condition that the expression under the logarithm in \eqref{eq-disc:entropywithisland-without-boundary} be positive (which is equivalent to the condition of spacelike separation of the points $\ba_{+}$ and $\bb_{+}$ or $\ba_{-}$ and $\bb_{-}$), which leads to $t_a < t_b +r_*(r_b)-r_*(r_a)$, it follows that $t_a +t_b >  2r_{*}(r_0)-r_{*}(r_a)-r_*(r_b)$, which means that $t_a >  r_{*}(r_0)-r_{*}(r_a)$ corresponds to the late time regime \eqref{eq-disc:late-time-regim-condition}.}
    \be
        \begin{aligned}
        \cosh \kappa_h (t_a + t_b) &>  \cosh \kappa_h (2r_{*} (r_0) - r_{*} (r_{a})-r_{*} (r_{b})), \\ 
        \cosh 2 \kappa_h t_{a} &< \cosh 2 \kappa_h (r_{*} (r_0) - r_{*} (r_{a})).
        \end{aligned}
        \label{eq-disc:intermediate-time-regim-condition}
    \ee
    It can be shown that in the intermediate time regime there is no solution for an island near the horizon (see Appendix \ref{appendix-B} for details).

    \item \underline{Late time regime}:

    This time regime is determined by the following condition
    \be
        \begin{aligned}
        \cosh \kappa_h (t_a + t_b) &\gg  \cosh \kappa_h (2r_{*} (r_0) - r_{*} (r_{a})-r_{*} (r_{b})), \\ 
        \cosh 2 \kappa_h t_{a} &\gg \cosh 2 \kappa_h (r_{*} (r_0) - r_{*} (r_{a})).
        \end{aligned}
        \label{eq-disc:late-time-regim-condition}
    \ee
    Under this assumptions one can also explicitly calculate  the approximate location of the horizon 
    \bea\label{eq:late-time-ta}
    t_a &\approx& t_b - \frac{1}{2 \kappa_h} \log\left[1-\frac{27 \pi^3}{c^3 G^3 \kappa^6_h} e^{2\kappa_h (r_*(r_0)+2r_*(r_b)-t_b)-3}\right], \\
    r_a &\approx& r_h+\frac{c^2 G^2 e^{(r_h-r_b)/r_h}}{144 \pi^2 r^2_h (r_b-r_h)}-\frac{12 \pi r_h(r_0-r_h)(r_b-r_h)e^{\frac{r_0+r_b-2r_h-t_b}{r_h}}}{c G}. \label{eq:late-time-ra}
    \eea
    The formula \eqref{eq:late-time-ta} must be real-valued, so this implies the estimate for the existence of the island solution
    \be\label{eq:late-time-condition-that-island-exist}
    t_b > t^2_b, \qquad t^2_b = r_{*}(r_0)+2r_*(r_b)-\frac{3}{2\kappa_h} \log \left( \frac{c G \kappa^2_h e}{3\pi}\right).
    \ee
     In the limit $r_0 \to \infty$ \eqref{eq:late-time-condition-that-island-exist}  one can see that  $t^2_b \rightarrow \infty$, i.e. this means absence of the late time regime  as it should be.
\end{itemize}

\hspace{0pt}

\begin{figure}[t!]\centering
    \includegraphics[width=0.4687\textwidth]{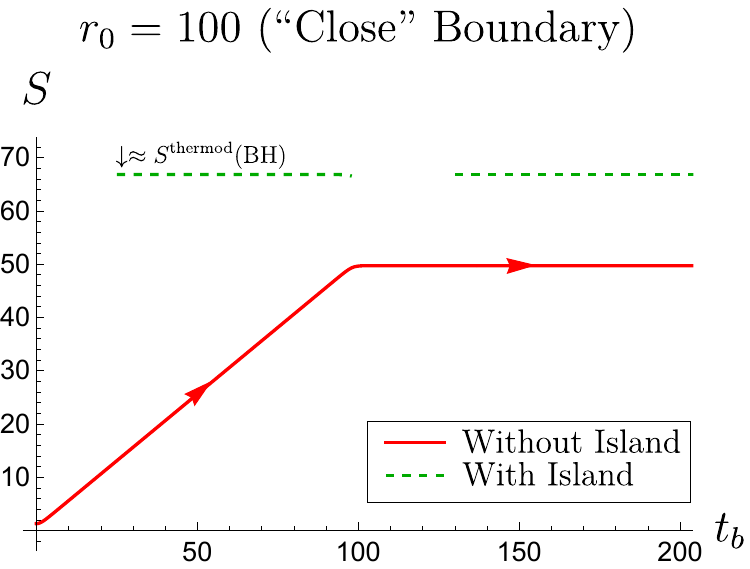}
    \qquad \includegraphics[width=0.4687\textwidth]{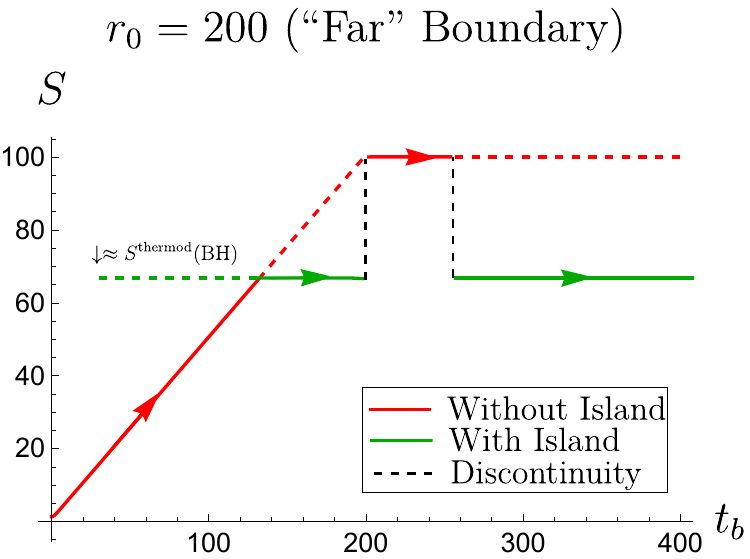}
	\caption{Time dependence of entanglement entropy taking into account nontrivial island configuration for a black hole with symmetric boundary for different boundary locations $r_0$. }
 \label{fig:entropy-with-island-discc}
\end{figure}

The difference between the times $t^2_b$ and $t^1_b$ (see \eqref{eq:late-time-condition-that-island-exist} and \eqref{eq:early-time-condition-that-island-exist}, respectively), that we call ``{\it blink time}'' $t_{\text{blink}}$, is equal to
\be
\begin{aligned}
&t_{\text{blink}} \equiv t^2_b-t^1_b = 3 r_* (r_b)-\frac{3}{2\kappa_h} \log \left( \frac{c G \kappa^2_h e}{3\pi}\right)= \\ &= 6 G M \left( \frac{r_b}{2 G M} - \log \left[\frac{c e}{48 \pi G M^2 (r_b/2GM - 1)} \right]\right).
    \end{aligned}
\label{eq:difference-of-times-existence}
\ee
for a fixed ratio $r_b/2 G M = \text{fixed}$ and the fixed number of fields $c = \text{fixed}$, with an increase in the mass of the black hole~$M$ the time difference $t_{\text{blink}}$ increases (as we mentioned earlier, calculations are carried out assuming $c G \kappa^2_h \ll 1$).
 From \eqref{eq:difference-of-times-existence} this condition leads to the crucial relation
\be\label{eq:time-when-island-does-not-exist}
t^2_b > t^1_b.
\ee
It implies that during the time interval $t_b \in (t^1_b, t^2_b)$ island configuration near horizon does not exist. In the other words the general picture of  island evolution is the following
\begin{itemize}
    \item  First the island appears and exists for some time disappearing at  $t_b = t^1_b$ 
    \item  At time  $t_b \in (t^1_b, t^2_b)$ island solution near horizon does not exist
    \item After the time $t^2_b$  it appears again  and stays until the infinite time
\end{itemize}

Formally for a large number of fields $c$ it follows from \eqref{eq:difference-of-times-existence} that $t^2_b < t^1_b$ for
\be\label{eq:very-big-nubmer-of-fields}
c > \frac{3 \pi}{G \kappa^2_h} e^{2\kappa_h r_* (r_b)-1}.
\ee
So for such $c$ the island does not disappear for a finite period of time. However, we emphasize that \eqref{eq:very-big-nubmer-of-fields} violates our assumption that $c G \kappa^2_h \ll 1$.

\begin{figure}[t!]\centering
    \includegraphics[width=0.4687\textwidth]{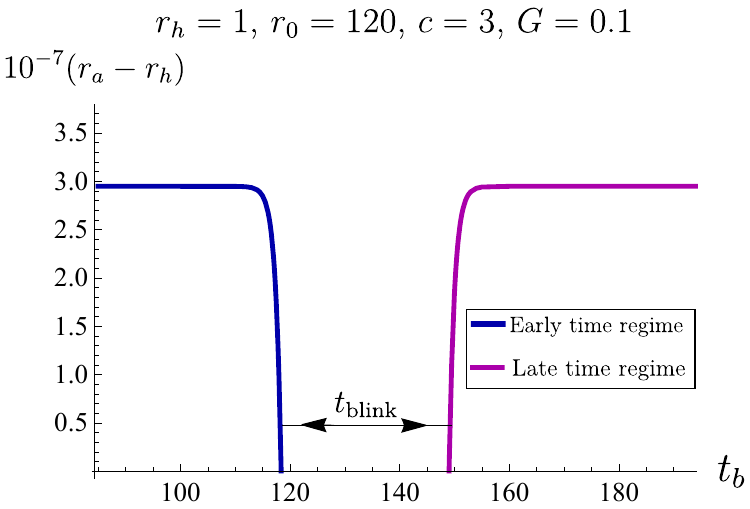}
    \qquad \includegraphics[width=0.4687\textwidth]{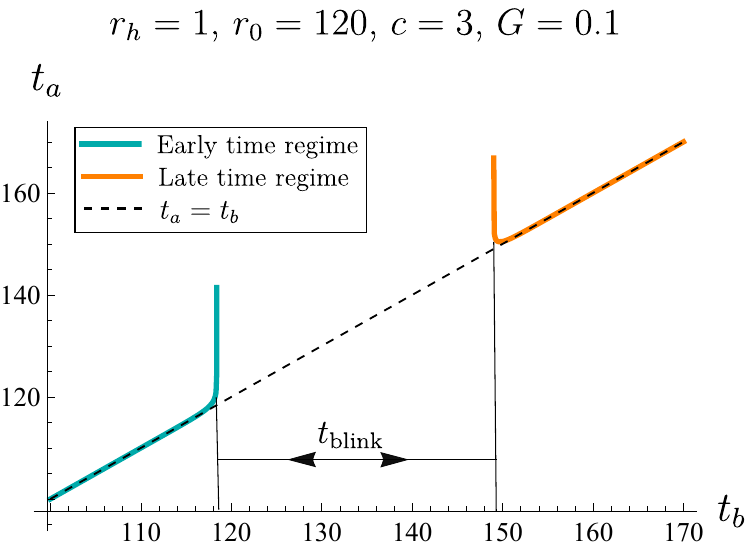}
	\caption{Dependence on the time $t_b$ of the island location  $r_a$ (\textbf{left})/time $t_a$ (\textbf{right}).}
    \label{fig:solution-extremization-disc}
\end{figure}
Thus the effect of the boundary in general is that it leads to the possibility of emergence what we call ``blink effect'' -- for a short time the island protecting us from the unitarity violation is switched off, thus leading to a short-time unitarity violation. Let us  summarize in details the behaviour of entropy and the influence of this new effect of its dynamics
\begin{itemize}
    \item As it was said in the previous subsection, for a boundary ``close enough'' to the event horizon, the entropy without an island reaches saturation before it exceeds the thermodynamic entropy of the black hole and the entropy with the island (see Fig.\ref{fig:entropy-with-island-discc} left). Thus, entanglement entropy without an island dominates throughout evolution, and non-unitary evolution does not take place, because condition \eqref{eq:condition-of-unitarity} is satisfied for all times.

    \item For a ``far enough'' boundary, for which the entropy without an island at some moment exceeds the thermodynamic entropy of a black hole, the situation is different. In this case, the island configuration, at least at some times, removes monotonous growth, which is consistent with unitary evolution. However, due to the disappearance of the island for a finite period of time $t_{\text{blink}}$ \eqref{eq:difference-of-times-existence}, the entanglement entropy jumps from a constant value of entropy with the island to a constant value of entropy without the island at saturation \eqref{eq:disc-saturation-value} (see Fig.\ref{fig:entropy-with-island-discc} right). Since the latter is greater than the thermodynamic entropy of a black hole, non-unitary evolution takes place, that is
    \be
    S(R_2) (t_b) > S^{\text{therm}}_{BH}, \qquad t_b \in (t^1_b, t^2_b).
    \ee

    \item Summarizing we conclude that for a ``close enough'' boundary, an island is not needed for unitary evolution, and for a ``far enough'' boundary the unavoidable disappearance of island for a finite period of time $t_{\text{blink}}$ leads to non-unitary evolution.

    \item From the calculational viewpoint this could be explained from the behaviour of extremization equations solutions, i.e. the  dependence on time $t_b$  island location $(r_a, t_a)$, (see Fig.\ref{fig:solution-extremization-disc} demonstrating the numerical solution and \eqref{eq:early-time-ta}, \eqref{eq:early-time-ra}, \eqref{eq:late-time-ta}, \eqref{eq:late-time-ra} for approximate analytical solutions). Most of the time during which the island exists, these solutions correspond to the ``canonical" case without boundaries~\cite{Hashimoto:2020cas}. However, in the vicinity of the moments of time $t^1_b$ (time $t^2_b$) when the island disappears (appears), the radial coordinate $r_a$ decreases (increases) significantly, approaching the horizon $r_h$ (moving away from the horizon $r_h$), while the time coordinate $t_a$ increases (decreases) significantly.
    
\end{itemize}
\subsection{Single-sided region}\label{sec:Single-sided region}
\begin{figure}[h!]\centering
    \includegraphics[width=0.95\textwidth]{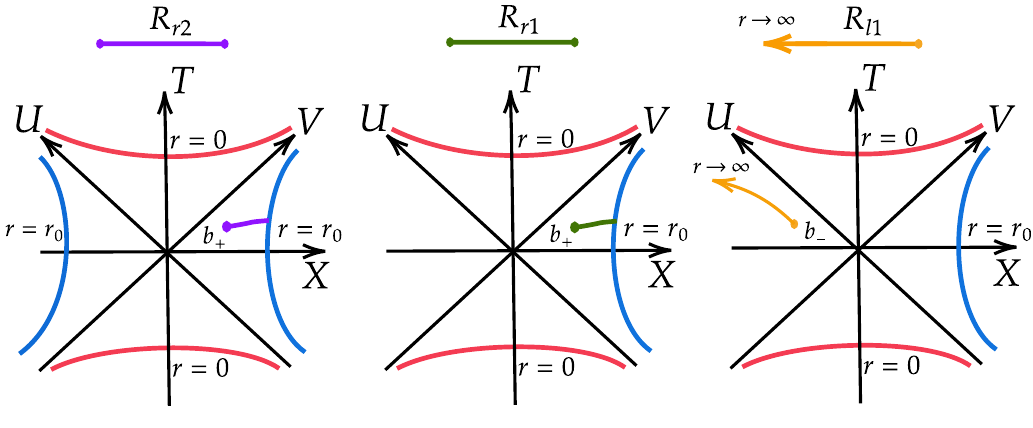}
	\caption{The schematic depicture of three types of one-wedge region. \textbf{Left}: region $R_{r2}$ (magenta) in double-boundary Lorentzian geometry. \textbf{Center}: region $R_{r1}$ (green) in single-boundary geometry similar to $R_{r2}$. \textbf{Right}: semi-infinite region $R_{l2}$ (orange) in left wedge of single-boundary geometry.}\label{fig:one-point-region-plots}
\end{figure}

Let us finalize this section with consideration of the entanglement region $R_{r2}$ without entanglement dynamics in the double-boundary geometry. Let us choose this  region~
$R_{r2}$ being located only in the right wedge  between spatial points $r_b$ and $r_0$   at some time $t_b$, see Fig.\ref{fig:one-point-region-plots} (left). The entanglement entropy of $R_{r2}$ in two-boundary geometry is just the constant and has no time dependence
\be\label{eq-disc:entropyonepoint}
S(R_{r2}) = \frac{c}{6} \log\left( \frac{2 \sqrt{f(r_b)}\sinh \kappa_h (r_{*}(r_0)-r_{*}(r_b))}{\kappa_h \eps} \right).
\ee
It is curious that a similar region $R_{r1}$ for a single-boundary geometry, see Fig.\ref{fig:one-point-region-plots} (center), which will be considered in the next  section has the dynamics.
Note that the entropy~\eqref{eq-disc:entropyonepoint} infrared diverges when the boundary is removed, i.e. at $r_0 \to \infty$.

\section{Single-boundary geometry}\label{sec:single}
 \begin{figure}[h!]\centering
    \includegraphics[width=0.915\textwidth]{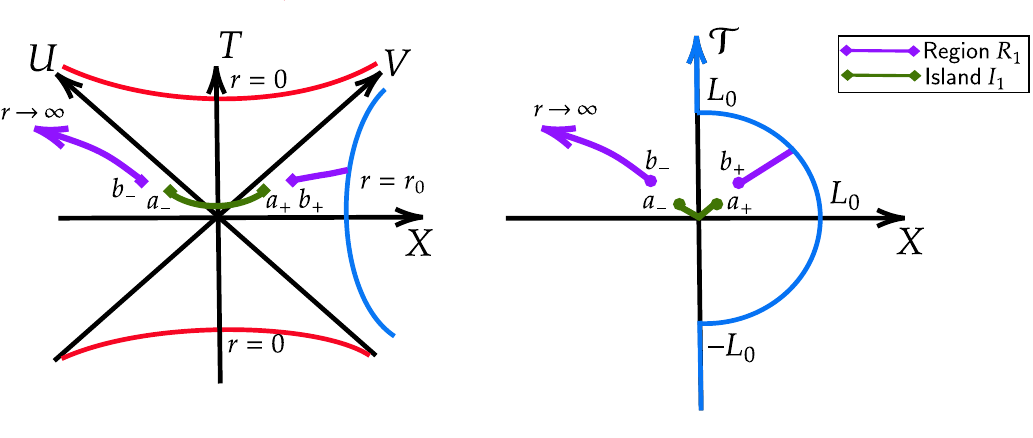}
	\caption{The schematic depicture of the region $R_1$ (magenta) in: two-boundary Lorentzian geometry (\textbf{left}) and its Euclidean version (\textbf{right}).} \label{fig:region-symm}
\end{figure}
\subsection{Without island}
Let us start the study of entanglement entropy in single-boundary geometry introducing the analog of region $R_2$  (denote it $R_1$) and calculating the entanglement entropy $S(R_1)$ for it. The region $R_1$ in the right wedge coincides with $R_2$ being restricted by the boundary location $r_0$ while in the left wedge which has no boundary it extends from point $\bb_-$ to spatial infinity, see Fig. \ref{fig:region-symm}. As in the previous section we first calculate the entanglement entropy $S(R_1)$ without inclusion of island mechanism 
\be\label{eq:odna-zaslonka-two-point-no-island}
    \begin{aligned}
&S(R_1) = \frac{c}{6} \log \left( \frac{4 f(r_b) \cosh^2 \kappa_h t_b}{\kappa^2_h \eps^2}  \right) + \frac{c}{6} \log \left( \frac{\cosh 2\kappa_h t_b +\cosh 2\kappa_h (r_{*}(r_0)-r_{*}(r_b))}{\cosh^2 \kappa_h t_b}\right)\\ &+ \frac{c}{6} \log \left( \frac{3}{8} \sin \left[\frac{2}{3} \arctan \frac{\cosh \kappa_h t_b}{\sinh \kappa_h (r_{*}(r_0)-r_{*}(r_b))}\right] \sin \left[ 2 \arctan \frac{\cosh \kappa_h t_b}{\sinh \kappa_h (r_{*}(r_0)-r_{*}(r_b))}\right]\right). 
    \end{aligned}
\ee
Note that in the limit $r_0 \to \infty$ the last two terms of \eqref{eq:odna-zaslonka-two-point-no-island} tend to zero, so the entropy without boundary (the first term) is restored. 

\begin{figure}[t!]\centering
    \includegraphics[width=0.55\textwidth]{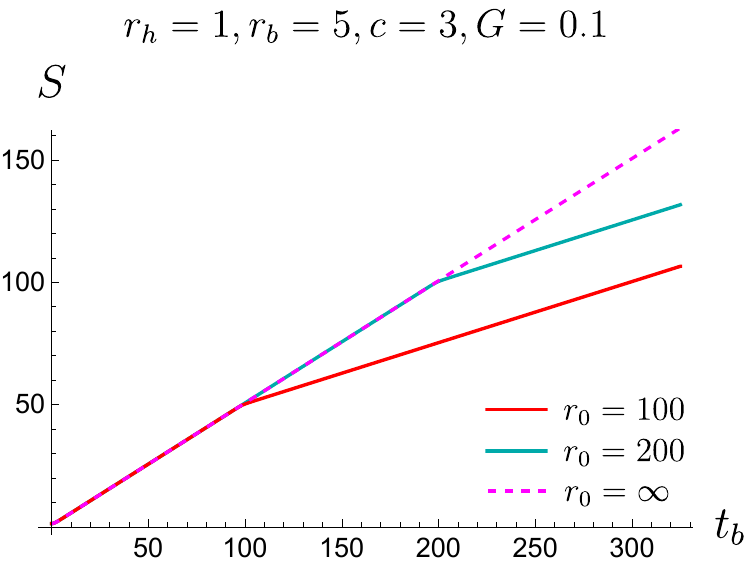}
	\caption{Time-dependence of  entanglement entropy without the  island term  for the region $R_1$.} \label{fig:one-b-noisland}
\end{figure}

The exclusion of one of the boundaries lead to the following picture of entanglement entropy dynamics (see Fig.\ref{fig:one-b-noisland}): first the system follows the linear growth regime $S(R_1)\approx c\kappa_h t_b/3$ exactly as in the previous cases, then
 approximately at time $t^1_b = r_* (r_0) -r_* (r_b)$ the entropy \eqref{eq:odna-zaslonka-two-point-no-island} reduces to the twice weaker linear growth
\bea\label{eq:odna-zaslonka-collector}
S(R_1) \simeq \frac{c \kappa_h t_b}{6}+ \frac{c}{6} \log \left( \frac{3 \sqrt{3} f(r_b)}{2\kappa^2_h \eps^2} \sinh \kappa_h (r_{*}(r_0)-r_{*}(r_b)) \right). 
\eea
Note that for a similar region $R_2$, in the case of two symmetric boundaries, the entropy without an island~\eqref{eq-disc:entropynoisland1} reaches a constant \eqref{eq:disc-saturation-value} at approximately the same time $t^1_b$. This difference can be interpreted so that in the case of one boundary, ``equilibrium'' occurs only in one wedge, while in the case of two boundaries -- in both wedges.

We also note one more important difference. For the case of two boundaries, the entropy without an island \eqref{eq-disc:entropynoisland1} at a sufficiently close location of the boundary $r_0$ relative to the horizon does not exceed the thermodynamic entropy of the black hole~\eqref{eq:thermodynamic-entropy-of-black-hole}. For the single-boundary case, the entropy without an island \eqref{eq:odna-zaslonka-two-point-no-island}, due to the presence of monotonous growth at all times, exceeds the thermodynamic entropy \eqref{eq:thermodynamic-entropy-of-black-hole} for any location of the boundary $r_0$. Thus, if there is no island configuration for such a region, then a non-unitary evolution takes place for all $r_0$.

\subsection{Blinking island in single-boundary geometry}

\begin{figure}[t!]\centering
    \includegraphics[width=0.4687\textwidth]{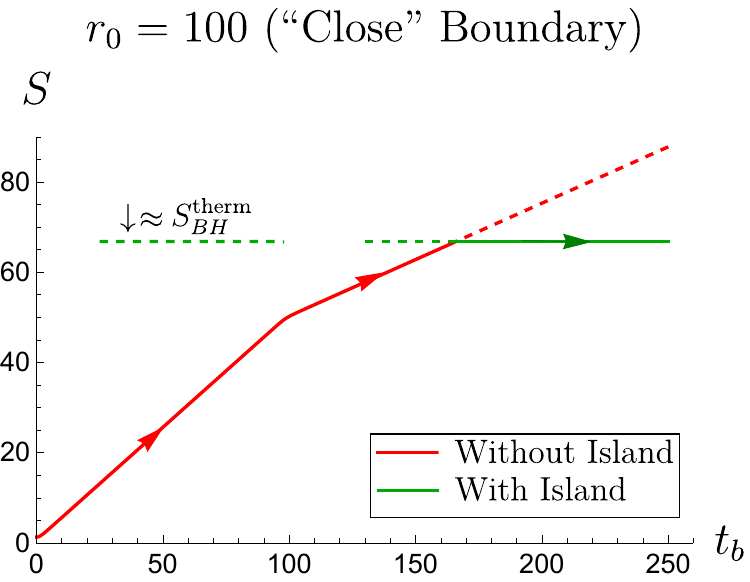}
    \qquad \includegraphics[width=0.4687\textwidth]{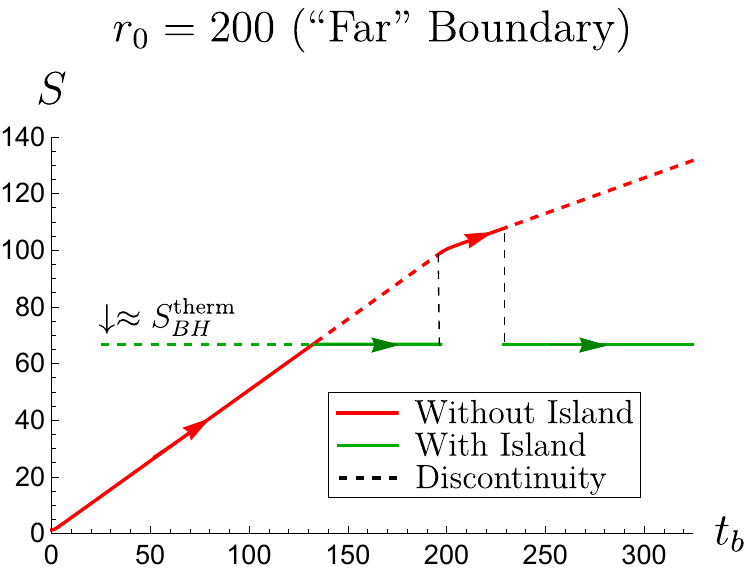}
	\caption{Time dependence of entanglement entropy taking into account nontrivial island configuration for a black hole with only one boundary for different boundary locations $r_0$. }
 \label{fig:entropy-with-island-disc}
\end{figure}

So far in the previous section we observe that entanglement entropy in the  geometry with asymmetric boundaries also lead us to the unbounded growth.  In this subsection for region $R_1$ we consider the entanglement entropy with a non-trivial island $I_1$ and examine the possibility whether the island mechanism saves unitarity as well in the symmetry  or one could observe ``blinking island''. Again  we assume that to avoid non-unitarity issues entanglement entropy should not  exceed thermodynamic entropy of black hole  for all times
\be\label{eq:condition-of-unitarity-odna}
S(R_1) (t_b) < S^{\text{therm}}_{BH}
\ee
where the thermodynamic entropy of a black hole is also given by the \eqref{eq:thermodynamic-entropy-of-black-hole}.

The presence of asymmetry between the right and left wedges indicates that the island configuration $I_1$ does not have to be symmetrical (with $\bb_\pm$ given by \eqref{eq:points-for-our-region}). Therefore, we select the island endpoints as
\be
\ba_+ = \left(r_{a_+}, t_{a_+} \right), \qquad \ba_- = \left(r_{a_-}, -t_{a_-} \right),
\ee
where the point $\ba_{+}$ ($\ba_{-}$) is located in the right (left) wedge.
The generalized entropy~\eqref{eq:gen_functional} with the island $I_1$ is
\be
S_\gen[I_1, R_1] = \frac{\pi r^2_{a_+}}{G}+\frac{\pi r^2_{a_-}}{G}+S_\m(R_1 \cup I_1),
\label{eq-disc:entropyWithisland1}
\ee
where the matter entanglement entropy $S_\m(R_1 \cup I_1)$ is calculated by applying the formulas \eqref{eq:entropy-for-Dirac-fermions-UHP}-\eqref{eq:weyl}  for region $R_1 \cup I_1$. Unfortunately, in contrast to the case with two symmetrical boundaries, we could not find a readable analytical formula for $S_\m(R_1 \cup~I_1)$ due to the difficult conformal transformation \eqref{eq:conform-transformation-one-boundary} for single-boundary geometry. All calculations for entanglement entropy with an island (i.e. numerical solution of extremization equation and use of island formula) in this section are performed  numerically. As a result we find blinking island solution in single-boundary geometry (for region~$R_1$). The time-dependence of entanglement entropy including the blinking island effect is presented in Fig.\ref{fig:entropy-with-island-disc} and it is qualitatively identical to  the double-boundary geometry case (except the presence of linear growth with different coefficients at different time scales).

\subsection{Single-sided regions}

\subsubsection*{Exterior with boundary}
Finally let us consider the dynamics of entanglement entropy for regions located in the single exterior. As we have shown in section \ref{sec:Single-sided region}, in double-boundary geometry region completely located in a single exterior is time-independent. In contrast to this we find that in the single-boundary geometry  such type of region now has the dynamics. Again, to set up notation, we consider the geometry with the right exterior containing the boundary, and left is free of it. We denote the region in right exterior as $R_{r1}$ -- this region is finite and bounded by some point $\bb_+$ and the boundary, see Fig.\ref{fig:one-point-region-plots} (center).
\begin{figure}[t!]\centering
    \includegraphics[width=0.44\textwidth]{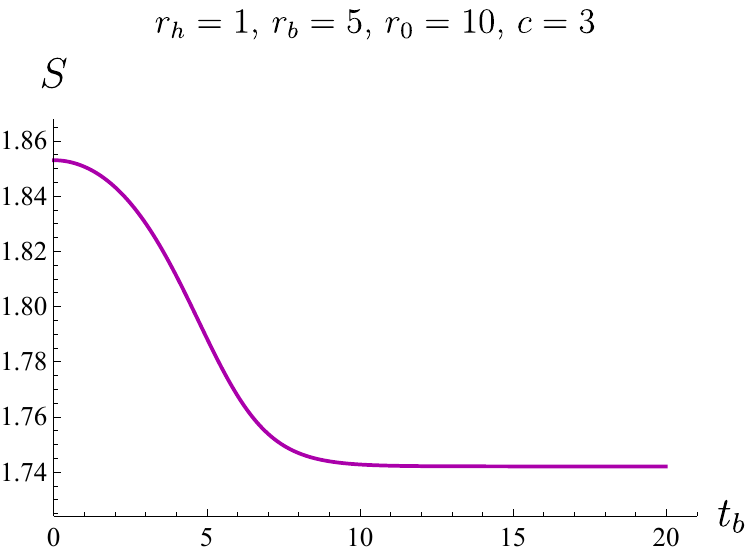} \includegraphics[width=0.44\textwidth]{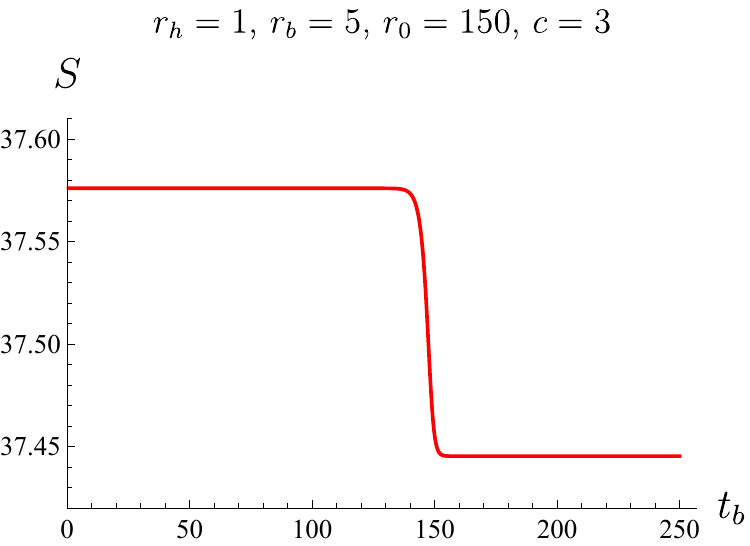}
	\caption{The time-dependence for entanglement entropy $S(R_{r1})$ for different reflecting boundaries values $r_0$.}\label{fig:decr-ent}
\end{figure}
Removing one of the boundaries leads to the dynamics of entanglement entropy. Explicitly,  it is given by 
\be
    \begin{aligned}
S(R_{r1}) =&  \, \frac{c}{12} \log \left( \frac{9 f(r_b)}{2\kappa^2_h \eps^2 } \left[\cosh 2\kappa_h t_b+\cosh 2\kappa_h (r_* (r_0)-r_*(r_b))\right] \right)\\ \, &+\frac{c}{6}\log \cos \left( \frac{\pi}{6}+\frac{2}{3}\arctan \frac{\cosh \kappa_h t_b}{\sinh \kappa_h (r_*(r_0)-r_*(r_b))}\right),
    \end{aligned}
\ee
and approximately at time $t^1_b = r_* (r_0) -r_* (r_b)$ it decrease to 
\be
S(R_{r1}) \simeq \frac{c}{6} \log\left( \frac{2 \sqrt{f(r_b)}\sinh \kappa_h (r_{*}(r_0)-r_{*}(r_b))}{\kappa_h \eps} \right).
\ee
which coincides with \eqref{eq-disc:entropyonepoint}. From Fig.\ref{fig:decr-ent} one can see that the entanglement entropy is the constant until some time moment defined by boundary location just decreases to some constant value (for larger $r_0$  this transition is more sharp). If the boundary $r_0$ is relatively close to the horizon $r_h$ then  the evolution already starts with the decrease.

\begin{figure}[t!]\centering
    \includegraphics[width=0.45\textwidth]{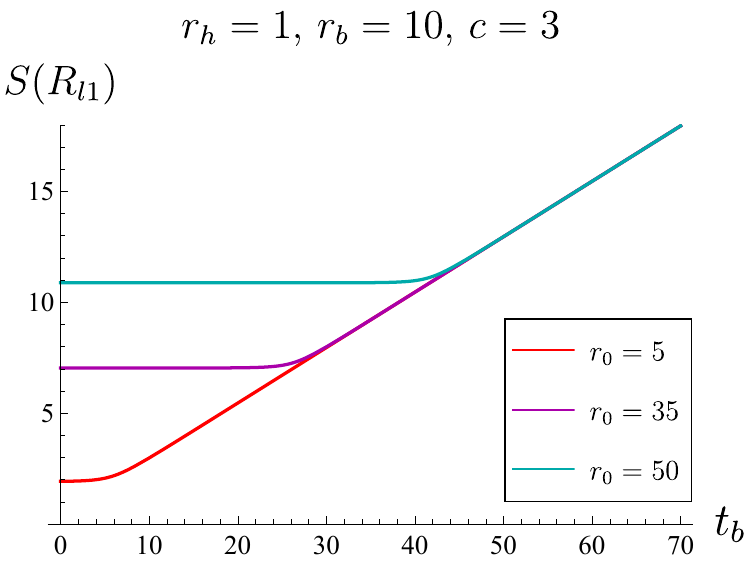}
 \label{fig:left-region-no-island}
 \caption{Time dependence of entanglement entropy \eqref{eq:left-region-no-island-one-boundary} without an island for the region $R_{l1}$ located in the left wedge, for geometry with a boundary only in the right wedge at different values of $r_0$.}
\end{figure}

\subsubsection*{Exterior without boundary}

By analogy let us consider the region $R_{l1}$  -- region between spatial infinity and some points $\bb_-$ in the left wedge, see Fig.\ref{fig:one-point-region-plots} (right). Again the entanglement entropy $S(R_{l1})$ is time-dependent and is given by
\be\label{eq:left-region-no-island-one-boundary}
    \begin{aligned}
S(R_{l1}) =&  \, \frac{c}{12} \log \left(\frac{9 f(r_b)}{2 \kappa^2_h \eps^2} \left[ \cosh 2 \kappa_h(r_* (r_0) -r_*(r_b)) +\cosh 2 \kappa_h t_b \right] \right) \\ \, &+\frac{c}{6} \log  \sin  \left(\frac{\pi}{3} \pm \frac{2}{3} \arctan \frac{\cosh \kappa_h t_b}{\sinh \kappa_h (r_* (r_0) -r_* (r_b))} \right),
    \end{aligned}
\ee
where the plus (minus) corresponds to $r_0 > r_b$ ($r_0 < r_b$). Remind that for double-boundary geometry or geometry without boundaries such a region has no dependence on time. Now if in the ``entangled universe on the other side'' one confined black hole in the box we will see the linear growth for $S(R_{l1})$ approximately at $t_b = |r_* (r_0) -r_* (r_b)|$
\bea\label{eq:odna-zaslonka-two-point-no-island-slow-growth}
S(R_{l1}) \simeq \frac{c \kappa_h t_b}{6}+ \frac{c}{6} \log \left( \frac{3 \sqrt{3} f(r_b)}{4\kappa_h \eps}\right).
\eea
Closer boundary ``on the other side'' is located to the horizon -- faster the growth will start ``on this side''. For this case one can find the asymmetric insland solution numerically (for simplicity we do not present them).
\section{Discussion and main results}\label{sec:DMR}
Let us briefly summarize our results. We deform the geometry of eternal black hole introducing reflecting  boundaries  at some distance from the horizon. We consider two situations, double-boundary geometry, when we place boundaries symmetrical in each exterior, and single-boundary geometry, when only one exterior contains boundary.

$\,$

For these types of geometries we find that:
\begin{itemize}
\item In contrast to geometry without any boundary which has the unlimited entanglement entropy growth (when entangling region is semi-infinite and located in both exteriors), in double-boundary geometry the growth stops after some time proportional to the boundary location. When location of the boundary tends to infinity, $r_0\rightarrow \infty$, this time also becomes infinite. In this way for some $r_0$ we do not have information paradox at all -- the entropy never exceeds thermodynamical entropy. In some sense this resembles the situation with calculation of heat capacity -- without boundaries the capacity of Schwarzschild  solution is negative which indicates the instability. However, if we include the boundaries then the capacity becomes positive for some black hole masses \cite{York:1986it}. 

\item  Examining asymmetric geometry where the boundary is present in only one exterior (i.e. total state corresponds to two entangled states, one with the boundary and one without it), we observe that entanglement entropy becomes dynamical for entanglement regions that are completely contained within the single exterior. Remind that when left and right exteriors are symmetric in the same situation no dynamics in the entanglement entropy could be observed.
\end{itemize}

It would be interesting to clarify whether this effect is just the drawback of such setup involving s-wave approximation or manifestation of  possible problems concerning entanglement islands in higher-dimensional gravity, see discussion in \cite{Geng:2021hlu,Geng:2023iqd,Geng:2023qwm}.

\begin{acknowledgments}

The work of IA and TR, which consisted of calculating the entanglement entropy in the context of entanglement islands for the case of symmetric reflecting boundaries (Chapters 1-3 of this paper) is supported by the Russian Science Foundation (project 24-11-00039, Steklov Mathematical Institute). TR also participated in the discussion of comparison of results obtained in Chapter 3 and Chapter 4. The work of DA, consisting in calculating the time evolution of the entanglement entropy in the presence of one reflecting boundary, is supported by the Foundation for the Advancement of Theoretical Physics and Mathematics “BASIS” (Chapter 4). Also the work of DA was performed at Steklov International Mathematical Center and supported (Chapter 4) by the Ministry of Science and Higher Education of the Russian Federation (Agreement No. 075-15-2022-265).

\end{acknowledgments}

\newpage

\appendix
\section{Conformal map from single-boundary geometry to UHP}\label{appendix-A}

In this subsection, we prove that the conformal transformation from the Euclidean geometry of a two-sided black hole with a boundary in only one wedge (see Fig.\ref{appendix-conformal-map-figure}.A), associated with the complex coordinate $w = X+ i {\cal T}$ \eqref{eq:complex-coordinate-omega}, to the upper half-plane is given by \eqref{eq:conform-transformation-one-boundary}.

As discussed in the setup, the domain under consideration in Euclidean single-boundary geometry in the plane $({\cal T}, X)$  corresponds to the union of left half-plane $X<0$ (denote it $\Omega_1$) and the interior of a semi-disk in the right half-plane $X>0$ with the radius $L_0 = e^{\kappa_h r_{*} (r_0)}/\kappa_h$ (denote it $\Omega_2$). So, the domain in question of the complex plane $\omega$ is $\Omega = \Omega_1 \cup \Omega_2$. Let us express the complex variable $w$ in exponential form
\be
w = \rho e^{i \theta}.
\ee
We present the conformal map \eqref{eq:conform-transformation-one-boundary} from $\Omega$ to UHP as a composition of several simple conformal maps: $\Omega \to \Sigma \to H \to \text{UHP}$, see Fig.\ref{appendix-conformal-map-figure}.
\begin{itemize}
    \item Transform $\Omega \to \Sigma$ by inversion with multiplication by constant
    \be\label{app:first-map}
        \sigma = \frac{i L_0}{w}, \quad \sigma = \varrho e^{i \varphi} \, \, \, \Rightarrow \, \, \, \varrho = \frac{L_0}{\rho}, \, \,  \varphi = \frac{\pi}{2}-\theta.
    \ee
    The domain $\Omega_1 = \{\rho \in [0, +\infty), \, \, \theta \in \left[\pi/2, 3\pi/2\right]\}$ under conformal map \eqref{app:first-map} transforms to the lower half-plane $\Sigma_1 = \{\varrho \in [0, +\infty)$, $\varphi \in \left[-\pi, 0 \right]\}$.

    The domain $\Omega_2 = \{\rho \in [0, L_0], \, \, \theta \in \left[-\pi/2, \pi/2\right]\}$ under conformal map \eqref{app:first-map} transforms to the exterior of unit semi-disk in the upper half-plane \\
    $\Sigma_2 = \{\varrho \in [1, +\infty)$, $\varphi \in \left[0, \pi\right]\}$.

    The resulting domain of the complex plane $\sigma$ is $\Sigma = \Sigma_1 \cup \Sigma_2$ (see Fig.\ref{appendix-conformal-map-figure}.B).

    \item Transform $\Sigma \to H$ by fractional-linear transformation
    \be\label{app:second-map}
        \eta = \frac{\sigma-1}{\sigma+1}, \,\,\, \eta = \xi e^{i \lambda} \Rightarrow \xi = \sqrt{\frac{1+\varrho^2-2\varrho \cos \varphi}{1+\varrho^2+2\varrho \cos \varphi}}, \, \, \, \lambda = \arctan \left( \frac{2 \varrho \sin \varphi}{\varrho^2 -1} \right)  + k \pi.
    \ee
    where $k=0$ ($k=1$) is for $\varrho > 1$ ($\varrho < 1$).

    The domain $\Sigma_1$ under conformal map \eqref{app:second-map} transforms to the lower half-plane $H_1 = \{\xi \in [0, +\infty)$, $\lambda \in \left[-\pi, 0 \right]\}$.

   The domain $\Sigma_2$ under conformal map \eqref{app:second-map} transforms to the first quadrant $H_2 = \{\xi \in [0, +\infty)$, $\lambda \in \left[0, \pi/2\right]\}$.

    The resulting domain of the complex plane $\eta$ is $H = H_1 \cup H_2$ (see Fig.\ref{appendix-conformal-map-figure}.C).

\begin{figure}[t!]\centering
    \includegraphics[width=0.95\textwidth]{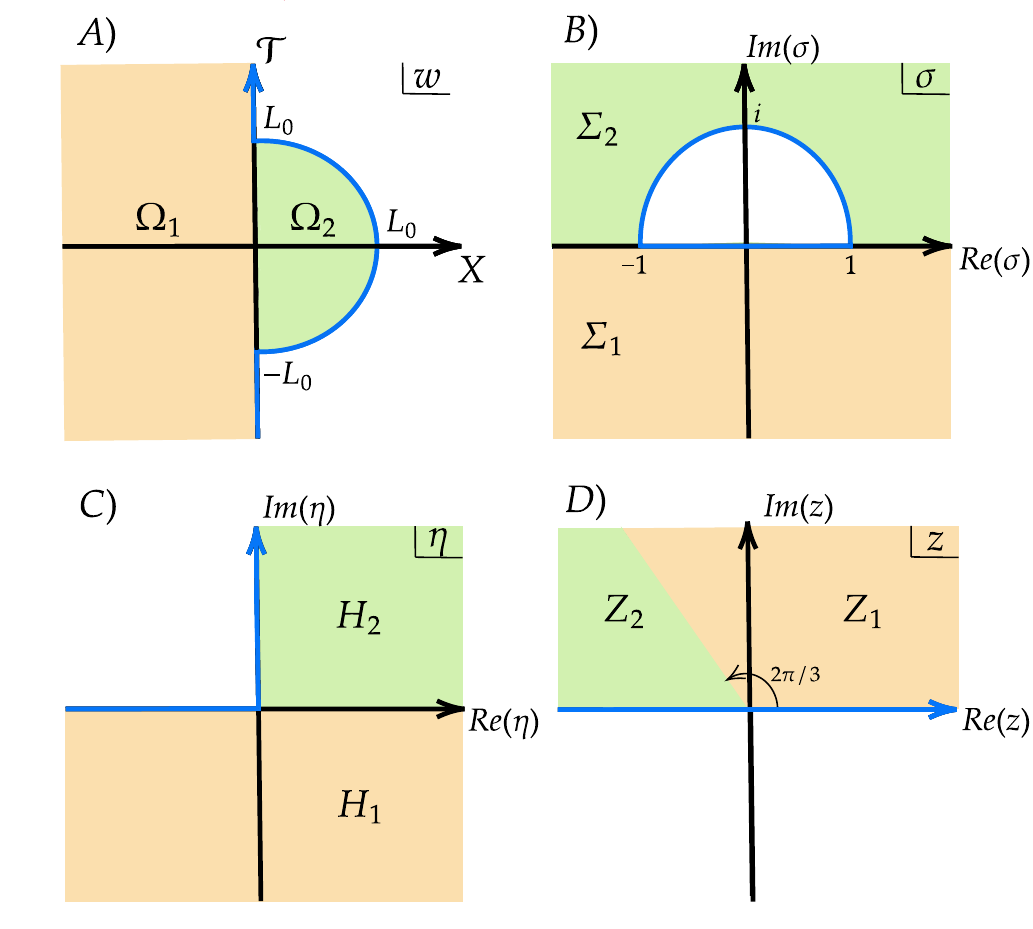}
 \caption{A sequence of conformal transformations from the union of the left half-plane and the interior of the unit disk in the right half-plane (A) to the upper half-plane~(D).}
  \label{appendix-conformal-map-figure}
\end{figure}

    \item Transform $H \to \text{UHP}$ by
    \be\label{app:third-map}
        z = e^{2\pi i/3} \eta^{2/3}, \, \, \,  z = \chi e^{i \psi} \quad \Rightarrow \quad \chi = \xi^{2/3}, \, \, \, \psi = \frac{2}{3} \left( \lambda + \pi \right),
    \ee
    where we choose the root of the third degree of a complex number $\zeta = x \, e^{i y}$ that corresponds to the principal branch $\zeta^{1/3} = x^{1/3} e^{i y/3}$.

    The domain $H_1$ under conformal map \eqref{app:third-map} transforms to the sector in the upper half-plane $Z_1 = \{\chi \in [0, +\infty)$, $\psi \in \left[0, 2\pi/3\right]\}$.

    The domain $H_2$ under conformal map \eqref{app:third-map} transforms to the sector in the upper half-plane $Z_2 = \{\chi \in [0, +\infty)$, $\psi \in \left[2\pi/3, \pi\right]\}$.

    The resulting domain of the complex plane $z$ is $\text{UHP} = Z_1 \cup Z_2$ (see Fig.\ref{appendix-conformal-map-figure}.D).
    
\end{itemize}
Combining the above conformal maps \eqref{app:first-map}--\eqref{app:third-map}, one can obtain the conformal transformation $\Omega \to \text{UHP}$ given by \eqref{eq:conform-transformation-one-boundary}.

\section{The absence of island in the intermediate time regime}\label{appendix-B}

To obtain the entanglement entropy, it is necessary to carry out the extremization of the generalized entropy with respect to coordinates $(r_a, t_a)$ of the island configuration endpoints $\ba_{\pm}$, see formula \eqref{eq:extremization-system-disc} and discussions about it.

In this appendix, we explicitly show that for the case with symmetrical boundaries in both exteriors, considered in Section \ref{sec:symm}, an island near horizon, i.e. with $r_a = r_h + X$, $X/r_h \ll 1$, does not exist in the intermediate time regime \eqref{eq-disc:intermediate-time-regim-condition} for $c G \kappa^2_h \ll 1$.

For the convenience of analytical calculations, instead of extremization by $(r_a, t_a)$, we consider extremization by $\left(r_*(r_a), t_a \right)$. More precisely, we show that under the above conditions, the following system of equations
\begin{equation}\label{app:extremization-system-disc}
 \begin{cases}
   \displaystyle\frac{\partial S_\gen[I_2, R_2]}{\partial t_a} = 0, 
   \\
    \displaystyle\frac{\partial S_\gen[I_2, R_2]}{\partial r_*(r_a)}- \displaystyle\frac{\partial S_\gen[I_2, R_2]}{\partial t_a} = 0.
 \end{cases}
\end{equation}
where $S_\gen[I_2, R_2]$ is given by \eqref{eq:points-for-symmetric-island}, does not have the required solution near horizon.
\\

We perform calculations under the following natural assumptions
\be\label{app:conditions}
\begin{aligned}
& r_*(r_a) < 0, \qquad e^{2\kappa_h r_*(r_a)} \ll 1, \qquad \kappa_h (r_*(r_0)-r_*(r_a)) \gg 1,\\ & \kappa_h t_{a,b} \gg 1, \qquad 2 r_*(r_0)-r_*(r_a)-r_*(r_b) \gg |t_a - t_b|.
    \end{aligned}
\ee
Let us denote $S_\gen[I_2, R_2]={\cal S}$. Under the conditions \eqref{app:conditions} the explicit expression for the time derivative is
\be\label{app:time-derivative-full}
\begin{aligned}
& \frac{\partial {\cal S}}{\partial t_a}   \simeq  \frac{c \kappa_h}{3} \Big(\frac{2\sinh \kappa_h (t_a-t_b)}{e^{\kappa_h(2r_*(r_0)-r_*(r_a)-r_*(r_b))}}-\frac{e^{\kappa_h (t_a+t_b)}}{e^{\kappa_h (r_* (r_b)-r_* (r_a))}+e^{\kappa_h (t_a+t_b)}} - \frac{e^{2\kappa_h t_a}}{e^{2\kappa_h (r_* (r_0)-r_* (r_a))}+e^{2\kappa_h t_a}} \\ & -\frac{2 \sinh \kappa_h (t_a-t_b)}{e^{\kappa_h (r_*(r_b)-r_*(r_a))}-2\cosh \kappa_h(t_a-t_b)}+\frac{e^{\kappa_h (t_a+t_b)}}{e^{\kappa_h (2r_*(r_0)-r_* (r_b)-r_* (r_a))}+e^{\kappa_h (t_a+t_b)}}+1\Big),
    \end{aligned}
\ee
and an explicit expression for the difference of the derivatives is given by
\be\label{app:difference-between-derivatives}
\begin{aligned}
\frac{\partial {\cal S}}{\partial r_*(r_a)} - \frac{\partial {\cal S}}{\partial t_a}  \simeq \, & \frac{2\pi e^{2\kappa_h r_*(r_a)-1}}{G \kappa_h} + \frac{c \kappa_h}{3} \Big( 1 - \frac{e^{\kappa_h (r_*(r_b)-r_*(r_a))}-2\sinh \kappa_h (t_a-t_b)}{e^{\kappa_h (r_*(r_b)-r_*(r_a))}-2\cosh \kappa_h (t_a-t_b)} \\ &-2 e^{2\kappa_h r_*(r_a)-1}+2 e^{-\kappa_h (2r_*(r_0)-r_*(r_a)-r_*(r_b)+t_a-t_b)}\Big).
    \end{aligned}
\ee
Let us simplify the time derivative \eqref{app:time-derivative-full} in accordance with the intermediate time regime \eqref{eq-disc:intermediate-time-regim-condition}, considering that condition \eqref{eq:time-for-existense-of-island-in-early-regim} is satisfied. Then
\bea\label{app:time-derivative-intermediate-regime}
\frac{\partial {\cal S}}{\partial t_a} \simeq \frac{c \kappa_h}{3} \Big(1+\gamma   -\frac{2 \sinh \kappa_h (t_a-t_b)}{e^{\kappa_h (r_*(r_b)-r_*(r_a))}-2\cosh \kappa_h(t_a-t_b)} \Big),
\eea
where
\be
\begin{aligned}
\gamma = & -\frac{e^{\kappa_h (2 r_* (r_0)-r_*(r_a)-r_*(r_b))}}{e^{\kappa_h (t_a+t_b)}}-\frac{e^{2\kappa_h t_a}}{e^{2\kappa_h (r_*(r_0)-r_*(r_a))}} \\ &+ \frac{e^{\kappa_h (r_*(r_b)-r_*(r_a))}}{e^{\kappa_h (t_a+t_b)}} +\frac{2\sinh \kappa_h (t_a-t_b)}{e^{\kappa_h (2r_*(r_0)-r_*(r_a)-r_*(r_b))}} \ll 1.
    \end{aligned}
\ee
Equating the expression on the right-hand side of \eqref{app:time-derivative-intermediate-regime} to zero, we obtain
\bea\label{app:solutions-full}
e^{\kappa_h (t_a-t_b)} = \frac{e^{\kappa_h (r_*(r_b)-r_*(r_a))}(1+\gamma) \pm \sqrt{e^{2\kappa_h (r_*(r_b)-r_*(r_a))}(1+\gamma)^2-4\gamma(2+\gamma)}}{2(2+\gamma)}.
\eea
We expand these expressions with respect to small $\gamma \ll 1$
\be\label{app:solutions-simple}
e^{\kappa_h (t_a-t_b)} \simeq\left\{\begin{array}{cc}
   \left(1+  \gamma/2\right) e^{\kappa_h (r_*(r_b)-r_*(r_a))}/2, \\
    \gamma  \,  e^{-\kappa_h (r_*(r_b)-r_*(r_a))},
\end{array}\right.
\ee
where the top (bottom) row corresponds to the plus (minus) in \eqref{app:solutions-full}.

Substituting the top row from \eqref{app:solutions-simple} to the difference \eqref{app:difference-between-derivatives}, we get
\bea\label{app:difference-for-upper-solution}
\frac{\partial {\cal S}}{\partial r_*(r_a)} - \frac{\partial {\cal S}}{\partial t_a} \Big|_{\text{top row}}  \simeq 2 c \kappa_h e^{2 \kappa_h r_*(r_a)-1} \left[\frac{\pi}{c G \kappa^2_h} - \frac{1}{3} \left( 1+4e^{1-2\kappa_h r_*(r_b)} + \ldots \right) \right] ,
\eea
where the ellipsis denotes terms are even smaller than those indicated in parentheses. Note that $4 e^{1-2\kappa_h r_*(r_b)} \ll 1$ due to $r_h \ll r_b$. So, since the calculations are carried out in the approximation $c G \kappa^2_h \ll 1$, it turns out that the first term in~\eqref{app:difference-for-upper-solution}, which is positive, is much larger than the negative second term. Therefore, the difference \eqref{app:difference-for-upper-solution} is positive and does not equal to zero.

Substituting the bottom row from \eqref{app:solutions-simple} to the difference \eqref{app:difference-between-derivatives}, we get
\bea\label{app:difference-for-lower-solution}
\frac{\partial {\cal S}}{\partial r_*(r_a)} - \frac{\partial {\cal S}}{\partial t_a}  \Big|_{\text{b. row}} \simeq 2 c \kappa_h e^{2 \kappa_h r_*(r_a)-1} \left[\frac{\pi}{c G \kappa^2_h} + \frac{1}{3} \left( e^{1-2\kappa_h r_*(r_a)}-1 + \ldots \right) \right].
\eea
Note that $e^{1-2\kappa_h r_*(r_a)} > 1$ due to $r_*(r_a) < 0$. So, it is clear that the expression \eqref{app:difference-for-lower-solution} is positive and essentially different from zero.

Thus, it is shown that in the intermediate time regime under natural conditions there is no nontrivial island configuration such that its endpoints are near the horizon.

\end{document}